%% `Cordelia.tex': SM as a noncommutative geometry, low energy regime
%%   by C.~P.~Martin, J.~M.~Gracia-Bondia and J.~C.~Varilly
%% (Plain TeX format, some AMS fonts needed)

\magnification\magstep1
\scrollmode
\overfullrule=0pt
\hfuzz=1pt

\font\tenbbb=msbm10 \font\sevenbbb=msbm7         %% extra AMS fonts
\font\tenams=msam10 \font\sevenams=msam7 \font\tenbm=cmmib10
\font\sevenbm=cmmib10 at 7pt \font\eightit=cmti8 %% for the addresses
\font\eightrm=cmr8 \font\eighti=cmmi8            %% for the abstract
\font\eightsy=cmsy8 \font\eightams=msam8

\newfam\bbbfam \newfam\amsfam \newfam\bmfam      %% (\fams 8,9,A)
\textfont\bbbfam=\tenbbb \scriptfont\bbbfam=\sevenbbb
\def\Bbb{\fam\bbbfam}                            %% Blackboard bold
\textfont\amsfam=\tenams
\scriptfont\amsfam=\sevenams  %% AMS symbols
\textfont\bmfam=\tenbm \scriptfont\bmfam=\sevenbm
\def\bm{\fam\bmfam}                              %% Bold math letters

\def\opname#1{\mathop{\rm#1}\nolimits} %% operator name

\def\a{\alpha}                      %% abbreviation for \alpha
\def\A{{\Bbb A}}	                %% noncommutative gauge potential
\def\Al{{\cal A}}                   %% an algebra
\def\Aut{\opname{Aut}}              %% automorphism group
\def\b{\beta}                       %% abbreviation for \beta
\def\Bar#1{\overline{#1}}           %% abbreviation for \overline
 %% graded tensor product
          %% bra vector <w|
\def\C{{\Bbb C}}                    %% complex numbers
\def\cite#1{{\rm[#1]}}              %% reference citation
\def\col{{\rm col}}                 %% colour algebra label
\def\D{{\cal D}}                    %% Dirac operator for real K-cycle
\def\d{{\rm d}}                     %% commutative differential
\def\del{\partial}                  %% abbreviation for \partial
\def\delslash{{\partial\mkern-9mu/}}%% standard Dirac operator
\def\diag{\opname{diag}}            %% diagonal matrix
\def\Diff{\opname{Diff}}            %% diffeomorphism group
                    %% a projective module
\def\eps{\epsilon}                  %% abbreviation for \epsilon
\def\eq#1{{\rm(#1)}}                %% print equation number
\def\F{{\Bbb F}}                    %% noncommutative gauge field
\def\ga{\gamma}                     %% abbreviation for \gamma
\def\GeV{\,\opname{GeV}}            %% particle mass unit
\def\H{{\cal H}}                    %% a Hilbert space
\def\HH{{\Bbb H}}                   %% quaternions
                %% identity operator
   %% implication arrow
           %% items with separation
    %% itemitems with separation
                    %% finite rank operators
          %% ket vector |z>
 %% operator |z><w|
\def\L{{\cal L}}                    %% a Lagrangian
\def\La{\Lambda}                    %% exterior algebra
\def\la{\lambda}                    %% abbreviation for \lambda
\def\mbar{\overline{m}}             %% renormalized mass term
           %% omit one symbol
\def\MSbar{\overline{\rm MS}}       %% modified minimal subtraction
\def\mtop{m_{\rm top}}              %% top mass
                    %% nonnegative integers
\def\Om{\Omega}                     %% abbreviation for \Omega
                     %% abbreviation for \omega
\def\Onda#1{\widetilde{#1}}         %% abbreviation for \widetilde
               %% tensor product over \Al
\def\ox{\otimes}                    %% tensor product
    %% repeated tensor product
\def\Q{{\Bbb Q}}                    %% rational numbers
 %% end-proof
\def\R{{\Bbb R}}                    %% real numbers
\def\row#1#2#3{{#1}_{#2},\dots,{#1}_{#3}} %% list:~a_1,...,a_n
\def\sepword#1{\qquad\hbox{#1}\quad} %% well-spaced words
\def\set#1{\{\,#1\,\}}              %% set notation
                   %% a sphere
                %% homemade blank box
                   %% symmetrically normed ideal
\def\Str{\opname{Str}}              %% supertrace of operator
\def\stroke{\mathbin\vert}          %% (for `\braket' and such)
                    %% circle group
\def\tfrac#1#2{{\textstyle{#1\over#2}}} %% small fraction in display
\def\thalf{\tfrac12}                %% small fraction 1/2
\def\tihalf{\tfrac i2}              %% small fraction i/2
\def\tiquarter{\tfrac i4}           %% small fraction i/4
\def\tquarter{\tfrac14}             %% small fraction 1/4
\def\Tr{\opname{Tr}}                %% trace of operator
\def\tr{\opname{tr}}                %% trace of matrix
\def\Trest{T\v{r}e\v{s}\v{t}}       %% a beautiful Czech castle
\def\Trs{\Tr^+}                     %% Dixmier's trace
\def\tthird{\tfrac13}               %% small fraction 1/3
\def\U{{\cal U}}                    %% unitaries
\def\vbar{\bar v}                   %% Higgs VEV
\def\w{\wedge}                      %% exterior product
           %% abbreviation; usage: \wrt~
\def\x{\times}                      %% products of various kinds
\def\Z{{\Bbb Z}}                    %% integers
\def\7{\dagger}                     %% adjoint of operator
\def\8{\bullet}                     %% anonymous degree
\def\.{\cdot}                       %% anonymous variable
\def\:{\colon}                      %% colon in  f:A->B
\def\<#1,#2>{\langle#1\stroke#2\rangle} %% Dirac brackets
\def\(#1,#2){(#1\stroke#2)}         %% another scalar product

%%% Special symbols:

\def\hexnumber#1{\ifcase#1 0\or 1\or 2\or 3\or 4\or 5\or 6\or 7\or
8\or 9\or A\or B\or C\or D\or E\or F\fi}
  %% fam8-->blackboard bold
\edef\theamsfam{\hexnumber\amsfam}  %% fam9-->AMS symbols
\edef\thebmfam{\hexnumber\bmfam}    %% famA-->bold math symbols

\mathchardef\gtrsim="3\theamsfam 26 %% greater-or-approx symbol
\mathchardef\lab= "7\thebmfam 15    %% Gell-Mann \lambda matrix
\mathchardef\lesssim="3\theamsfam 2E %% less-or-approx symbol
\mathchardef\taub= "7\thebmfam 1C   %% Pauli matrix

\newbox\ncintdbox \newbox\ncinttbox %% noncommutative integral symbol
\setbox0=\hbox{$-$}
\setbox2=\hbox{$\displaystyle\int$}
\setbox\ncintdbox=\hbox{\rlap{\hbox
to \wd2{\hskip-.125em\box2\relax\hfil}}\box0\kern.1em}
\setbox0=\hbox{$\vcenter{\hrule width 4pt}$}
\setbox2=\hbox{$\textstyle\int$}
\setbox\ncinttbox=\hbox{\rlap{\hbox
to \wd2{\hskip-.135em\box2\relax\hfil}}\box0\kern.1em}
\def\ncint{\mathop{\mathchoice{\copy\ncintdbox}{\copy\ncinttbox}
{\copy\ncinttbox}{\copy\ncinttbox}}\nolimits} %% \int-->\ncint

%%% Section divisions:

\newif\ifstartsec                   %% Flag for start of section

\outer\def\section#1{\vskip 0pt plus .15\vsize \penalty -250
\vskip 0pt plus -.15\vsize \bigskip \startsectrue
\message{#1}\centerline{\bf#1}\nobreak\noindent}

\def\subsection#1{\ifstartsec\medskip\else\bigskip\fi \startsecfalse
\noindent{\it#1}\penalty100\medskip}

\def\refno#1. #2\par{\vskip2pt\item{\rm\lbrack#1\rbrack}#2\par}%% Refs

\def\eightpoint{\normalbaselineskip=10pt %% for the abstract
\def\rm{\eightrm\fam0} \let\it\eightit
\textfont0=\eightrm \scriptfont0=\sevenrm %% (avoid using subscripts)
\textfont1=\eighti \scriptfont1=\seveni
\textfont2=\eightsy \scriptfont2=\sevensy \textfont\amsfam=\eightams
\normalbaselines \eightrm
\parindent=1em}

\hyphenation{geo-me-try}

%% Reference keys:

\def\ConnesEssay{1}
\def\ConnesLott{2}
\def\Book{3}
\def\MariaJose{4}
\def\ConnesGrav{5}
\def\Sirius{6}
\def\Blackone{7}
\def\Blacktwo{8}
\def\Orpheus{9}
\def\Persephone{10}
\def\Tresttalk{11}
\def\RealNCG{12}
\def\Proteus{13}
\def\ArtstateIV{14}
\def\SchZyl{15}
\def\IocSch{16}
\def\Neapolis{17}
\def\Testard{18}
\def\Chiron{19}
\def\Diracm{20}
\def\ChengLi{21}
\def\Donoghue{22}
\def\Nakahara{23}
\def\Witten{24}
\def\KaluzaK{25}
\def\CammaCoq{26}
\def\Peskin{27}
\def\Marshak{28}
\def\Belinda{29}
\def\ConnAction{30}
\def\Plassthesis{31}
\def\KPPW{32}
\def\CiprianiGS{33}
\def\KastlerEH{34}
\def\WolfMarkus{35}
\def\Connesphone{36}
\def\ChamCon{37}
\def\Kalau{38}
\def\CIKS{39}
\def\DoubleTrouble{40}
\def\Coque{41}
\def\Mainz{42}
\def\Fuzzy{43}
\def\Particledata{44}
\def\Onshell{45}
\def\MSbarra{46}
\def\Macha{47}
\def\Langacker{48}
\def\Veneziano{49}
\def\Kubo{50}
\def\CFFrohlich{51}
\def\Wulkenhaar{52}
\def\DuboisVKM{53}
\def\Sullivan{54}
\def\Becca{55}
\def\Pris{56}
\def\Geng{57}
\def\Ramond{58}
\def\LuisAG{59}
\def\Ramans{60}

%%% Document begins here:

\rightline{FT/UCM--12--96}
\rightline{UCR--FM--6--96}
\rightline{hep-th/9605001}

\vskip 1cm

%%% Title
\centerline{\bf The Standard Model as a noncommutative geometry:
the low energy regime}

\bigskip

%%% Authors
\centerline{\rm C. P. Mart{\'\i}n*}
\medskip
\centerline{\eightit Departamento de F{\'\i}sica Te\'orica I,
Universidad Complutense, 28040 Madrid, Spain}
\vfootnote*{email: {\tt carmelo@eucmos.sim.ucm.es}}
\bigskip
\centerline{\rm Jos\'e M. Gracia-Bond{\'\i}a and
Joseph C. V\'arilly\dag}
\medskip
\centerline{\eightit Departamento de Matem\'aticas, Universidad de
Costa Rica, 2060 San Pedro, Costa Rica}
\vfootnote\dag{email: {\tt varilly@cariari.ucr.ac.cr}}

\bigskip\bigskip

%%% Abstract
\begingroup\narrower\narrower
\eightpoint
We render a thorough, physicist's account of the formulation of the
Standard Model (SM) of particle physics within the framework of
noncommutative differential geo\-me\-try (NCG). We work in Minkowski
spacetime rather than in Euclidean space. We lay the stress on the
physical ideas both underlying and coming out of the noncommutative
derivation of the SM, while we provide the necessary mathematical
tools. Postdiction of most of the main characteristics of the SM is
shown within the NCG framework. This framework, plus standard
renormalization technique at the one-loop level, suggest that the
Higgs and top masses should verify
$1.3\, m_{\rm top} \lesssim m_H \lesssim 1.73\, m_{\rm top}$.
\par
\endgroup %% (end of abstract)

\bigskip

\section{1. Introduction}

It is seven years now since Connes and Lott conceived their
reconstruction of the Standard Model \cite{\ConnesEssay, \ConnesLott},
wrapped in a new branch of mathematics, noncommutative differential
geometry~\cite{\Book}.

The Connes--Lott approach restores a measure of symmetry and unity to
the SM, of itself regarded by many as an ugly {\it ad hoc\/}
contraption \cite{\MariaJose}. It gives naturally the correct
hypercharge assignments and indicates why the number of fermionic
families must be greater than one. In contrast to the conventional
version, the Higgs sector ---thus the weak vector boson mass matrix---
is at the output end of the model, and the properties of the
symmetry-breaking sector are determined.

In the Connes--Lott approach, external and internal degrees of freedom
of elementary particles are truly on the same footing. This is made
possible by the tools of NCG, that bridge the chasm between the
continuum and the discrete. Noncommutative geometry, in essence, is an
algebraic variational reformulation of the foundations of geometry,
extending to noncommutative spaces~\cite{\ConnesGrav}. Therefore, in
NCG the gravitational degrees of freedom are intrinsically coupled, at
the tree level, with the other fundamental interactions. The SM
Lagrangian corresponds to an effective theory in the limit of low
energies compared to the Planck mass. In that limit, fluctuations of
the noncommutative metric can be interpreted as noncommutative gauge
potentials, and the reduction to the gauge group of the SM is effected
by a unimodularity condition.

The present paper aims to deliver the tricks of the trade to the
working theoretical physicist. In some sense, it is a continuation of
\cite{\Sirius}, written in a more mathematical vein by two of us. Here
we concentrate on the substance of calculations, paying little heed to
mathematical niceties. We think the article worth the effort because:
(a) the NCG treatment of the SM has reached a degree of theoretical
maturity; (b) we have a better understanding now of the implications
of the NCG approach for model building in field theory; (c) there has
been new progress on the experimental front, such as the detection of
the top quark \cite{\Blackone, \Blacktwo}, with direct bearing on the
alleged predictions of the model; (d) the need we feel to amplify the
renormalization group analysis performed by two of us in
\cite{\Orpheus, \Persephone}; (e) the further need to discuss the
eventual falsification of Connes--Lott theory.

The version of Connes--Lott theory presented in this paper corresponds
to the ``new scheme'' \cite{\Tresttalk, \RealNCG} introduced by Connes
at the \Trest\ conference in May~1995. The older scheme, expounded in
\cite{\Book, \Sirius}, was criticized by Connes himself on the grounds
that the presence of a flavor algebra and a color algebra ---and
consequently of {\it two\/} noncommutative gauge potentials---
defeated the very purpose of unification inherent to the NCG approach.
We can add that a flavour-colour cross-term appears in principle, in
the two-algebra formalism, if neutrinos are massless~\cite{\Proteus}.

The noncommutative geometry derivation gives a constrained version of
the classical SM Lagrangian. That means that there are apparently
fewer arbitrary parameters than in the Glashow--Weinberg--Salam
formulation. Those restrictions have been seen as possible predictions
from NCG by some \cite{\ArtstateIV}. The most clear-cut of these
concerns the mass of the Higgs particle.

The paper is organized as follows. Section~2 has an heuristic
character: by looking at the fermionic sector of the SM Lagrangian, we
infer the particular noncommutative geometry associated to the
Standard Model. The concept of noncommutative gauge potential is made
explicit. In Section~3, we develop the main theoretical tools of
noncommutative geometry. Here we use Minkowskian $K$-cycles instead of
the Euclidean $K$-cycles favoured up to now.

Section~4 is a long computation, to wit, the step-by-step account of
the reconstruction of the bosonic sector of the SM, with a view to
determine which of the restrictions suggested by NCG on the parameters
of the Standard Model survive the use of the more general scalar
products allowed by the theory. We also deal with gauge reduction and
hypercharge assignments there. Section~5 is devoted to educated
guesses on the phenomenological status of those restrictions in
quantum theory.

It has been argued that ``most'' models of the Yang--Mills--Higgs type
cannot be obtained from noncommutative geometry \cite{\SchZyl,
\IocSch, \Neapolis}; it is then remarkable that the SM can. Thus, the
real usefulness, at present, of the NCG formulation is perhaps to
serve as a theoretical constraint on model building. In this respect,
the geometrical requirement of Poincar\'e duality is especially
strong. For instance, the SM in noncommutative geometry with
right-handed Dirac neutrinos thrown in does not satisfy all of the
conditions for Poincar\'e duality ---therefore does not correspond to
a noncommutative manifold in the strict sense \cite{\RealNCG,
\Testard}. Also, we have recently shown~\cite{\Chiron} that the
unimodularity condition is equivalent to cancellation of anomalies in
the NCG framework, for the algebra associated with the SM. All these
matters are taken up in Section~6.

\medskip

\section{2. Heuristic roots of noncommutative geometry}

Far from merely drawing a speculative picture of fundamental physics,
the NCG approach is solidly rooted in empirical fact. Its starting
point is a careful appraisal of the fermionic sector of the Lagrangian
of the Standard Model.

We quickly review the relevant facts. The elementary particles
discovered so far in Nature fall into two categories only: spin-one
bosons called gauge particles, and spin-$\thalf$ fermions. These two
categories are very different from both the physical and geometrical
standpoint. Physically, the spin-one bosons are the carriers of
fundamental interactions that come hand in hand with a gauge symmetry:
local $SU(3)$ for the strong force, and $SU(2) \x U(1)_Y$ for the
electroweak interactions. The physics of spin-$\thalf$ fermions is
obtained by applying Dirac's minimal coupling recipe \cite{\Diracm} to
the fermionic free action \cite{\ChengLi, \Donoghue}: fermions
interact by exchanging the gauge particles that their quantum numbers
allow.

The dynamics of the spin-one gauge bosons is governed by the
Yang--Mills action~\cite{\ChengLi, \Donoghue}. This action has a very
definite meaning in differential geometry: it is the square, in the
sense of inner-product, of the curvature of the gauge
connection~\cite{\Nakahara}. Geometrically, fermions are sections of
the spinor bundle. The minimally coupled Dirac operator, that is, the
Dirac operator twisted with a gauge connection, yields the gauge
covariant derivative of these sections.

Fermionic matter splits into left-handed and right-handed sectors.
Chirality or handedness, and the fact that $SU(2) \x U(1)_Y$ quantum
numbers of the left-handed fermions differ from the corresponding
quantum numbers of their right-handed partners, turn out to be crucial
ingredients of the NCG treatment.

In a chiral theory with an unbroken $SU(3) \x SU(2)_L \x U(1)_Y$
symmetry, gauge bosons and non-Majorana fermions such as electrons and
quarks will be massless. In nature, the $SU(2) \x U(1)_Y$ symmetry has
been broken down to $U(1)_{\rm em}$. The standard way of inducing such
a symmetry breaking implies including in the spectrum of the theory a
scalar particle, called the Higgs boson, by adding to the sum of
Yang--Mills and fermionic Lagrangians the so-called Higgs Lagrangian.
The latter has three parts: the gauge covariant kinetic term for the
Higgs field, the symmetry breaking Higgs potential, and the Yukawa
terms. Three of the original $SU(2) \x U(1)_Y$ gauge bosons become
massive via the Higgs mechanism. The Yukawa terms, in turn, render
massive the appropriate fermions.

However, the Higgs particle has not been detected yet, so there is no
definitive evidence that nature behaves in accordance with the
symmetry breaking pattern provided by the Higgs Lagrangian. The Higgs
sector of the SM Lagrangian is undeniably an {\it ad hoc\/} addition
to the original Lagrangian, its main virtue being that it provides a
both perturbatively renormalizable and unitary way of implementing the
breaking of the $SU(2)_L \x U(1)_Y$ symmetry. Nor does the Higgs
action receive any backing from the conventional geometrical side: it
lacks the beautiful geometrical interpretations of the Yang--Mills and
Dirac actions. Moreover, the Higgs particle is undeniably a very odd
particle in the Standard Model. It belongs to none of the particle
categories discovered so far, being the only scalar elementary
particle in the minimal Standard Model.

One may try to relate the Higgs to one of the aforementioned particle
categories. It obeys Bose--Einstein statistics, thereby standing
closer to gauge bosons than to fermionic matter. If we look at the
structure of interaction terms in the SM, we notice something
peculiar: the pointlike couplings between fermions and the Higgs
particle look very like those between fermions and gauge bosons.
Indeed, both gauge fields and the Higgs can be thought of as particles
that mediate interactions between fermions. The quartic
self{}interaction term is analogous to those of nonabelian Yang--Mills
fields. These are hardly original remarks, but noncommutative geometry
has the capacity to substantiate them, by reinterpreting the Higgs
field as a gauge field in a new sense. One would then expect Dirac's
minimal coupling recipe to yield at once the Yukawa terms and the
Dirac terms.

Of course, the Higgs will never be a gauge field within the framework
of classical differential geometry in four dimensions. Further (if
supersymmetry is not invoked) ordinary Kaluza--Klein field theories do
not lead to the SM due to the chiral structure of the fermionic
representations \cite{\Witten, \KaluzaK}.

\smallskip

Now we start to make the previous remarks concrete, in the spirit
of~\cite{\CammaCoq}. As advertised, we start by looking at the
fermionic sector of the SM. The fermion kinetic term for, say, the
first generation is written as follows:
$$
\eqalignno{
& \<\bar\ell_L, \ga^\mu D_\mu \ell_L>
+ \<\bar e_R, \ga^\mu D_\mu e_R> + \<\bar q_L, \ga^\mu D_\mu q_L>
+ \<\bar d_R, \ga^\mu D_\mu d_R> + \<\bar u_R, \ga^\mu D_\mu u_R>
\cr
&\qquad + \hbox{similar antimatter terms}.
& (2.1) \cr}
$$
Here $D_\mu \ell_L := i(\del_\mu - \tihalf Y(f) g_1 B_\mu
- \tihalf g_2\,\taub\cdot{\bm W}_\mu)\ell_L$ and
$D_\mu \ell_R := (\del_\mu - \tihalf Y(f) g_1 B_\mu)\ell_R$, where
$\ell$ denotes a chiral lepton; and also, if $q$ denotes a chiral
quark, $D_\mu q_L := (\del_\mu - \tihalf Y(f) g_1 B_\mu \allowbreak
- \tihalf g_2\, \taub\cdot{\bm W}_\mu
- \tihalf g_3\,\lab\cdot{\bm A}_\mu)q_L$ and
$D_\mu q_R := (\del_\mu - \tihalf Y(f) g_1 B_\mu
- \tihalf g_3\,\lab\cdot{\bm A}_\mu)q_R$; with $B$, $\bm W$ and
$\bm A$ denoting respectively the hypercharge, weak isospin and colour
gauge potentials and $Y(f)$ the various fermion hypercharges.

Let us collect the gauge potentials into a single package of
differential forms:
$$
\A' = i(B,W,A),
$$
where
$$
B = -\tihalf g_1 B_\mu \,\d x^\mu, \quad
W = -\tihalf g_2\,\taub\cdot{\bm W}_\mu \,\d x^\mu \quad{\rm and}\quad
A = -\tihalf g_3 \lab\cdot{\bm A}_\mu\,\d x^\mu.
$$
{}From the mathematical standpoint, $W$ is to be regarded as a 1-form
with values in the real field $\HH$ of quaternions. (In this paper,
ordinary differentials are denoted by an upright~$\d$.) In other
words, $\A'$ is an element of $\La^1(M) \ox \Al_F$, where
$\Al_F := \C \oplus \HH \oplus M_3(\C)$, that we baptize the
{\it Eigenschaften algebra}, is going to play a crucial r\^ole in the
NCG reconstruction of the SM, and we use $\La(M)$ to denote
differential forms on the Minkowski spacetime $M$.

Let us also collect all chiral fermion fields into a multiplet $\Psi$
and denote by $J$ the charge conjugation operation. Then \eq{2.1} is
rewritten as follows:
$$
I(\Psi) = \<\Psi, (i \delslash + \A' + J\A' J^\7) \Psi>.
\eqno (2.2)
$$
Next we look at the Yukawa part of the SM Lagrangian. It is convenient
to use the normalized Higgs doublet
$\Phi := \pmatrix{\Phi_1\cr \Phi_2\cr} := \sqrt2\,\phi/v$, where
$\phi$ is a true Higgs doublet with vacuum expectation
value~$v/\sqrt2$. We also need
$$
\Onda\Phi := \pmatrix{-\bar\Phi_2 \cr \bar\Phi_1 \cr}
= \pmatrix{0 &-1\cr 1 & 0\cr} \pmatrix{\bar\Phi_1 \cr \bar\Phi_2 \cr}.
$$

Just as the standard Dirac operator relates the left- and right-handed
spinor representations, the Yukawa operator $\D_F$ in the space of
internal degrees of freedom relates also the left- and right-handed
chiral sectors. The combination of both constructions yields a
Dirac--Yukawa operator (incorporating the Kobayashi--Maskawa
parameters) $\D = i\delslash \oplus \D_F$. Now, it has been remarked
by several people~\cite{\MariaJose, \CammaCoq} that the Higgs should
be properly regarded as a quaternion-valued field; this makes apparent
the custodial symmetry of the scalar sector of the SM, which is of
course related to the $\rho = 1$ tree level prediction for the
low-energy neutral to charged current interactions ratio. Introduce
then
$q_\Phi = \pmatrix{\bar\Phi_1 & \bar\Phi_2 \cr - \Phi_2 & \Phi_1 \cr}$
and $r := q_\Phi - \langle q_\Phi \rangle = q_\Phi - 1$ and write,
schematically for a right-left splitting of the fermion multiplets:
$$
\A'' = \pmatrix{& M^\7 r \cr r^\7 M & \cr},
$$
where $M$ denotes the quark or lepton masses, as the case might be.
Then the Yukawa terms for both particles and antiparticles (for the
first generation) are written
$$
\eqalign{
& I(\Psi,\A'',J) := \<\Psi, (\D_F + \A'' + J\A''J^\7) \Psi>
\cr
&= \bar q_L \Phi\,m_d\,d_R + \bar q_L \Onda\Phi\,m_u\,u_R
+ q_R \bar \Phi\,\bar m_d\, \bar d_L
+ q_R \Onda{\bar\Phi}\,\bar m_u\, \bar u_L
+ \bar\ell_L \Phi\,m_e\,e_R
+ \ell_R \bar\Phi\, \bar m_e\, \bar e_L + {\rm h.c.}
\cr}
$$
Note that $\Onda\Phi$ is absent in the lepton sector, since neutrinos
are massless, and that the hypercharges of $\Phi$, $\Onda\Phi$ are
respectively~$+1$ and ~$-1$.

\smallskip

With $\A := \A'\,\oplus\,\A''$, the {\it whole\/} fermionic sector of
the SM is thus recast in a form similar to~\eq{2.2}:
$$
I(\Psi) = \<\Psi, (\D + \A + J\A J^\7) \Psi>.
\eqno (2.3)
$$

\smallskip

In keeping with the old Kaluza--Klein idea, we have combined the
ordinary gauge fields and the Higgs as entries of a generalized gauge
field. The Yukawa terms come from the minimal coupling recipe applied
to the gauge field in the internal space. The Dirac--Yukawa operator
is seen to contain in NCG all the relevant information pertaining to
the~SM.

In the next Section, we show the mathematical consistency of the above
procedure, using the tools of noncommutative geometry as developed by
Connes~\cite{\Book}. The Connes--Lott approach ties in an intriguing
manner the properties of continuous spacetime with the intrinsic
discreteness stemming from the chiral structure of the SM. Left and
right-handed fermions are not to have the same $SU(2) \x U(1)_Y$
quantum numbers, and handedness and symmetry breaking by a Higgs
potential go together~\cite{\Witten}. Noncommutative geometry lends
support to the idea that ``the left and right-handed fermion fields
are fundamentally independent entities, mixed to form massive fermions
by some subsidiary process''~\cite{\Peskin}.

With the machinery of NCG in hand, we {\it derive\/} the bosonic part
of the Lagrangian in Section~4.

\section{3. $K$-cycles}

Modern particle dynamics is pervaded by the ``neutrino paradigm''
\cite{\Marshak}, that is to say, the principle of chirality
invariance. We regard the $K$-cycle, also called a spectral triple
---which, according to Connes, is the basic object of calculus on any,
in general noncommutative, space--- as the mathematical formalization
of the neutrino paradigm.

\smallskip

\subsection{3.1. Euclidean $K$-cycles and the noncommutative integral}

Practically all NCG computations till now have been regularized by use
of compact spaces with Euclidean metrics. In noncommutative geometry,
ordinary spaces are described by the algebras of functions defined
over them. In general, therefore, a compact noncommutative manifold is
just a ---not necessarily commutative--- selfconjugate algebra $\Al$
with a unit element~1. We shall suppose given a representation
of~$\Al$ on a Hilbert space $\H$. A compact $K$-cycle or generalized
Dirac operator on $\Al$ is a selfadjoint operator $D$ on~$\H$, with
compact resolvent, such that $[D,a]$ is bounded for all $a \in \Al$
and for some $d$ the operator $|D|^{-d}$ is integrable in the NCG
sense ---on which more below. An {\it even\/} $K$-cycle incorporates
in its definition a grading operator $\ga$ on $\H$; then $\ga$
commutes with the action of $\Al$ on $\H$ and anticommutes with~$D$.

There are two main prototypes for $K$-cycles. Consider first an
algebra which is the direct sum of two matrix subalgebras, ``right''
and ``left'': the simplest case is \hbox{$\C \oplus \C$}. Represent it
by several copies of the fundamental representation, and its
conjugate, on a finite dimensional Hilbert space. Take for $\ga$ the
operator equal to~1 on the right subrepresentation and to $-1$ on the
left one. The Dirac operator $D$ is then necessarily of the form
$$
D = \bordermatrix{&\scriptstyle R &\scriptstyle L \cr
\scriptstyle R & 0 & M^\7 \cr
\scriptstyle L & M & 0 \cr},
\eqno (3.1)
$$
that can be thought of as a matrix of Yukawa couplings. In this
example, the ``noncommutative integral'' is the ordinary trace.

Consider on the other hand a Riemannian space $M$ with spin structure.
Let $\H = L^2(S_M)$ be the Hilbert space of spinors $\psi$ on $M$.
Here $\Al$ is the space of functions $C^\infty(M;\C)$, represented as
a commutative multiplication algebra over $\H$. For (Clifford) actions
on spinors, of elements of a vector space of forms with orthonormal
basis $\{e^j\}$, we use the following notation: if
$v = v_j e^j = v_\mu\,\d x^\mu$, then
$\ga(v) := i \ga^j v_j = i \ga(\d x^\mu) v_\mu$, where the $\ga^j$
denote appropriate Dirac matrices. The Dirac operator $D$ is given by
$i\ga^j (e_j -\thalf \Gamma_{jk}^l \ga^k\ga_l)$, where the
$\Gamma_{jk}^l$ are components of the spin
connection~\cite{\Nakahara}, and all $K$-cycle specifications hold.

Moreover, the Dirac operator allows one in principle to recover the
metric from the following formula for the geodesic distance on~$M$:
$$
d(p,q) := \sup \set{|f(p) - f(q)| : f \in \Al,\ \|[D,f]\| \leq 1}.
\eqno (3.2)
$$
In physical terms: neutrinos constitute the simplest quantum probe for
classical geometry of a manifold~\cite{\Belinda}. A corresponding
formula from the kinematics of quantum scalar particles exists, but it
is much more complicated. Also, equation \eq{3.2} can be generalized
to turn the state space of any (nonnecessarily commutative) algebra,
endowed with a $K$-cycle, into a metric space.

In the present case the noncommutative integral is just the ordinary
integral over $M$ with its canonical volume element. If $M$ is of even
dimension, $\ga$ is the chirality operator $\ga_5$ and we have an even
$K$-cycle.

In general, the integral over a compact noncommutative space depends
on $D$ and is given by the Dixmier trace of compact operators on a
Hilbert space. We say that a positive compact operator $A$ with
eigenvalues $\la_1 \geq \la_2 \geq \cdots\,$ lies in the Dixmier trace
class if and only if $\la_1 + \cdots + \la_n = O(\log n)$. The Dixmier
trace is a generalized limit of the form
$$
\Trs A := \lim_{n\to\infty} {\la_1 +\cdots+ \la_n \over \log n},
$$
which can be extended to a linear functional on the class of operators
$A$ such that $\Trs \sqrt{A^\7 A}$ is finite. We say that a
noncommutative manifold $\Al$ is of dimension $d$ if
$\ncint 1 = \Trs |D|^{-d}$ exists and is nonzero.

When $\H = L^2(S_M)$ is the spinor bundle over a compact spin manifold
$M$ of dimension $d$ and $D$ is the standard Dirac operator, then
$|D|^{-d}$ lies in the Dixmier trace class. A fundamental trace
theorem of Connes \cite{\ConnAction} yields as a corollary, for any
$f \in C^\infty(M;\C)$, the following integral formula:
$$
\ncint f
:= \Trs f|D|^{-d} \propto \int_M f(x) \,\d^d\mathop{\rm vol}(x),
\eqno (3.3)
$$
where the proportionality constant depends only on~$d$.

In summary, the Dixmier trace, in the presence of a compact $K$-cycle,
gives a precise generalization for the integral over an ordinary
manifold. Chiral matrix algebras are \hbox{0-dimen}sional $K$-cycles;
then $\ncint$ reduces to the ordinary matrix trace. For more details
on $K$-cycles and the noncommutative integral, we refer
to~\cite{\Sirius}.

\subsection{3.2. Noncommutative differential calculus}

We need a noncommutative differential calculus as well. For a start,
one can embed $\Al$ in the ``universal differential algebra''
$\Omega\Al = \bigoplus_{n\geq0} \Omega^n\Al$, generated by symbols
$a_0\,da_1\dots da_n$ with a formal antiderivation $d$ satisfying
$d(a_0\,da_1\dots da_n) = da_0\,da_1\dots da_n$, $d\,1 = 0$ and
$d^2 = 0$. That algebra obviously is very large; but in the presence
of a $K$-cycle can be refined to a more useful one. This is done in
two steps. We first represent the whole of $\Om\Al$ on the Hilbert
space $\H$ by taking:
$$
\pi(a_0\,da_1\dots da_n) := a_0\,[D,a_1] \dots [D,a_n].
$$
The algebra of operators $\pi(\Om\Al)$ is not a differential algebra,
in general. This is because the kernel of $\pi$ is not closed under
$d$, as one can have $\pi(c) = 0$ with $\pi(dc) \neq 0$. The problem
is taken care of by a standard trick in algebra: the differential
ideal of ``junk''
$J := \{\,c' + dc'' \in \Om\Al: \pi c' = \pi c'' = 0\,\}$ is factored
out, thereby obtaining a new graded differential algebra of
``noncommutative differential forms'' $\Om_D\Al$ by
$$
\Om_D\Al := \Om\Al/J = \pi(\Om\Al)/\pi(J).
$$
The differential on $\Om_D\Al$, still denoted by $d$, is given by:
$$
d[\pi(a)] := [\pi(da)],
$$
where the classes $[\pi(\.)]$ are defined modulo $J$.

The astonishing fact, justifying the given name, is that this quotient
algebra, for the $K$-cycle of the continuum, is an algebra of
operators on $L_2(S_M)$ isomorphic to the de Rham algebra of
differential forms $\La(M)$. We outline the proof. The curvature terms
$\Gamma_{jk}^l$ of the Dirac operator on a compact space are
unimportant here, because they cancel out of the basic formula
$\pi(df) = [\delslash, f] = i\ga(\d f)$. If we think of the operators
$\ga(\d f)$ just as differential forms, we see that
$\Om_D^1 C^\infty(M,\C) \simeq \La^1(M,\C)$. Analogously,
$\pi(f_0\,df_1\,df_2) = - f_0 \ga(\d f_1) \ga(\d f_2)
= - f_0 \ga(\d f_1 \w \d f_2) - f_0 \d f_1\.\d f_2$. The last term is
junk: $\d f\.\d f = \pi(d (f\,df - \thalf df^2))$ and
$\pi(f\,df - \thalf df^2) = 0$. Therefore
$\Om_D^2 C^\infty(M,\C) \simeq \La^2(M,\C)$ and in general
$\Om_D^k C^\infty(M,\C) \simeq \La^k(M,\C)$. For more details, the
reader is again referred to~\cite{\Sirius}.

In the case of 0-dimensional $K$-cycles, the quotient algebra is a
{\it matrix differential algebra}. We exemplify the concept by working
out below, in full detail, the matrix differential algebras relevant
for the Standard Model.

This paper concerns itself with product $K$-cycles. It is well known
that the algebra of differential forms for the product of two ordinary
spaces is the skew tensor product of the two exterior algebras: this
is just the K\"unneth formula~\cite{\Nakahara}. To be more explicit,
the graded or skew tensor product tensor product $\Al_1 \ox \Al_2$ of
two graded associative algebras $\Al_1$, $\Al_2$ is just the tensor
product of the underlying vector spaces, endowed with a multiplicative
structure defined on elements of the form $a_1 \ox a_2$ by:
$$
(a_1 \ox a_2)(a'_1 \ox a'_2) = (-)^{\eps\eps'} a_1a'_1 \ox a_2a'_2,
$$
where $\eps$ is 0 or~1 according as $a_2$ is an even or odd element of
$\Al_2$, and similarly for $\eps'$ and $a'_1$. It has the obvious
graded structure.

We need the details of the construction of the algebra of
noncommutative differential forms for the tensor product of the
algebra of smooth functions on an even dimensional spin manifold $M$
and a matrix algebra $\Al_F$ represented in a finite dimensional
Hilbert space $\H_F$. In such a context, the product algebra of
noncommutative differential forms is the skew product of the de Rham
algebra and a matrix differential algebra, but the latter is not the
original matrix differential algebra $\Om_{D_F}\Al_F$, in general.
Nevertheless, in all cases of interest (i.e., when $\Al_F$ is a chiral
algebra with a nondegenerate $K$-cycle), the difference can only arise
at the level of second order and higher forms~\cite{\Plassthesis,
\KPPW}. Then we have $\Om^0_D\Al = \Al$ and $\Om^1_D\Al
= \La^1(M)\ox\Al_F\ \oplus\ C^\infty(M)\ox\Om^1_{D_F}\Al_F$.

\vfill\eject

\subsection{3.3. Noncommutative Yang--Mills actions}

The noncommutative integral allows one to define scalar products on
noncommutative differential forms. The prototype is
$$
\(S,T) := \ncint S^\7 T.
\eqno (3.4)
$$
A scalar product allows one to identify $\Om_D\Al$ with a subspace of
$\pi(\Om\Al)$ ---by choosing as representative of a class the
projection of any of its elements onto the subspace perpendi\-cular to
the junk ideal. It is understood that $\ncint$ is a trace, that is,
$\ncint AB = \ncint BA$ for all $A,B$. This holds for all well-behaved
$K$-cycles~\cite{\CiprianiGS}. With the scalar product \eq{3.4} and (a
multiple of) the usual scalar product of differential
forms~\cite{\Nakahara}, respectively, the isomorphism between
$\Om_D C^\infty(M)$ and $\La(M)$ becomes an isometry.

A connection $\nabla$ is, for our purposes, an antiderivation of
degree one on the graded algebra $\Om_D\Al$. The canonical connection
is given by the differential operator $d$; thus a connection must be
of the general form $\nabla = d + \A$, where $\A \in \Om_D^1\Al$
denotes the corresponding gauge potential in NCG. We ask $\nabla$ to
be compatible with the hermitian form $h$ on $\Al$ given by
$h(a_1,a_2) = a_1^\7 a_2$; that forces $\A$ to be selfadjoint. The
curvature of the connection is the noncommutative gauge field
$\F := \nabla^2 = d\A + \A^2$. The unitary group of the algebra
$\U(\Al) := \set{u \in \Al: uu^\7 = u^\7u = 1}$ defines gauge
transformations on the space of connections by
$\nabla \mapsto u\,\nabla\,u^\7$. Then it is easy to see that $\A$
transforms into ${}^u\A := u\,du^\7 + u \A u^\7$ and $\F$ maps to
${}^u\F := u\,\F\,u^\7$.

Note that conjugation by a unitary element is an (inner) automorphism
of the algebra $\Al$. The group of automorphisms $\Aut(\Al)$ of the
algebra is a noncommutative analogue of the group of diffeomorphisms
on an ordinary manifold. In fact, $\Aut(C^\infty(M)) \simeq \Diff(M)$.
On an ordinary manifold, conjugations by unitaries give trivial
automorphisms, but connections transform nontrivially.

The (noncommutative) Yang--Mills action $I_{YM}(\F)$ is given by the
quartic functional $\(\F,\F)$. In order to compute it, one requires
the precise description of the algebra $\Om_D\Al$ of noncommutative
differential forms up to degree two. Clearly, $I_{YM}(\F)$ is
invariant under gauge transformations. When applied to the $K$-cycle
of the continuum, the noncommutative Yang--Mills action falls back to
an ordinary Yang--Mills action. Thus, the notion of Yang--Mills action
in NCG directly extends the (bosonic part of the Euclidean version of)
quantum electrodynamics or quantum chromodynamics actions to the
context of noncommutative manifolds (but see the caveats at the end of
the next subsection).

\subsection{3.4. Real $K$-cycles}

The $K$-cycles of interest for us are {\it real\/} $K$-cycles. Besides
the quadruple $(\Al,\H,\D,\ga)$ that defines an even $K$-cycle, we
require that there be an antilinear isometry $J\: \H \to \H$, with the
following properties:
$$
\eqalignno{
&\quad J^2 = \pm1,  \qquad  J\D = \D J,  \qquad  J\ga = \pm\ga J;
& (3.5\,a) \cr
&[a, JbJ^\7] = 0,  \sepword{for all}  a,b \in \Al,
& (3.5\,b) \cr
&[[\D,a], JbJ^\7] = 0,  \sepword{for all}  a,b \in \Al.
& (3.5\,c) \cr}
$$
In most examples, in particular, for the Standard Model considered in
this paper, $\H$ doubles into a direct sum of eigenspaces $\H^+$,
$\H^-$ of particles and antiparticles. The physical role of $J$ is to
interchange them, therefore it is essentially a charge conjugation
operator. If $\xi \in \H^+$, then we write $\bar\xi := J\xi \in \H^-$;
similarly for $\bar\eta \in \H^-$. Then we have:
$$
\Psi = \pmatrix{\xi \cr \bar\eta} \in \H,  \sepword{with}
J \pmatrix{\xi \cr \bar\eta \cr} = \pmatrix{\eta \cr \bar\xi \cr}.
$$

Notice also that, on account of \eq{3.5$\,b$}, $\H$ is an bimodule over
the algebra $\Al$, with the associative operations given by
$$
\Psi \mapsto a \Psi b := a Jb^*J^\7 \Psi.
$$
This is perhaps a good place to fix our conventions on the different
involutions used in this paper: $^\7$ denotes the hermitian conjugate
of an operator; $^*$ shall denote the involutions on the several
differential algebras we shall introduce; we use $\bar\la$ for the
conjugate of a complex number $\la$.

The fermionic action for a real $K$-cycle is the functional \eq{2.3}:
$$
I(\Psi, \A, J) := \<\Psi, (\D + \A + J\A J^\7) \Psi>.
$$
The gauge group $\U(\Al)$ acts on the fermions in the adjoint
representation:
$$
\Psi \mapsto {}^u\Psi := u \Psi u^* = uJuJ^\7 \Psi.
$$
Even though the noncommutative fermionic multiplet $\Psi$ ``sees'' an
adjoint representation, the physical fermion fields transform in the
fundamental representation of the gauge group (or its conjugate). Note
that $\<\Psi, \Psi'> = \<{}^u\Psi,{}^u\Psi'>$, because both $u$ and
$J$ are isometries.

We have the following theorem:
$$
I(\Psi, \A, J) = I({}^u\Psi, {}^u\A, J).
$$
The proof needs the use of all of the properties of $J$ in~\eq{3.5}.
First we compute:
$$
\eqalign{
u JuJ^\7 \D Ju^*J^\7 u^*
&= JuJ^\7 u\D u^* Ju^*J^\7 = JuJ^\7 (\D + u[\D,u^*]) Ju^*J^\7
\cr
&= Ju\D u^*J^\7 + JuJ^\7 u[\D,u^*] Ju^*J^\7
\cr
&= J\D J^\7 + Ju[\D,u^*]J^\7 + u[\D,u^*]
\cr
&= \D + u[\D,u^*] + Ju[\D,u^*]J^\7.
\cr}
$$
Also, taking into account that $\A$ is of the form
$\sum_i a_i[\D,b_i]$, for $a_i,b_i \in \Al$, we get
$$
u JuJ^\7 \A Ju^*J^\7 u^* = u\A JuJ^\7 Ju^*J^\7 u^* = u\A u^*,
$$
and analogous manipulations give
$$
u JuJ^\7 (J\A J^\7) Ju^*J^\7 u^* = J u\A u^* J^\7.
$$
Therefore
$$
\eqalign{
\<{}^u\Psi, (\D + {}^u\A + J{}^u\A J^\7) {}^u\Psi>
&= \<uJuJ^\7 \Psi, uJuJ^\7 (\D + \A + J\A J^\7) Ju^*J^\7u^*uJuJ^\7\Psi>
\cr
&= I(\Psi, \A, J).
\cr}
$$

Our prototypes can be made into real $K$-cycles. Consider a finite
algebra $\Al_F$, as in subsection~3.1, chirally acting on the direct
sum of a Hilbert space and its conjugate. The Dirac operator $\D_F$
must now be of the form
$$
\D_F = \pmatrix{D_F & 0 \cr 0 & \Bar D_F \cr},
\eqno (3.6)
$$
where $D_F$ acting on $\H_F^+$ is given by \eq{3.1}, whereas $\Bar D_F$
is the conjugate matrix of Yukawa couplings. In this example we take
$J_F$ to be conjugation, including for convenience a chirality flip.

Connes' fundamental theorem of Riemannian geometry~\cite{\ConnesGrav}
establishes an intimate correspondence between real $K$-cycles and
Riemannian structures, in such a way that Riemannian geometry is given
a purely algebraic foundation. In Connes' axiomatics, the real
$K$-theory associated to the antilinear isometry $J$ plays a crucial
part in characterizing the type of spaces with the structure of (in
general, noncommutative) spin manifolds. In this way, charge
conjugation symmetry is given a geometrical r\^ole.

In order to see that, we paraphrase the foregoing proof in terms of
equivalence of $K$-cycles. Two $K$-cycles $(\Al,\H_1,\rho_1,\D_1)$ and
$(\Al,\H_2,\rho_2,\D_2)$ are equivalent when there is a unitary
operator $U$ such that $U \rho_1(a) U^* = \rho_2(a)$ and
$U \D_1 U^* = \D_2$, where we have explicitly written $\rho_1$,
$\rho_2$ for the representations of $\Al$ on $\H_1$ and $\H_2$. In the
context of real $K$-cycles, inner automorphisms of the algebra $\Al$
always give rise to such equivalences. To be more precise, the
automorphism $i_u(a) := u a u^*$, with $u \in \U(\Al)$, induces a
change of representation $\rho \mapsto i_u^* \rho:= \rho \circ i_u$.
Then we have just seen that the two real $K$-cycles $(\Al,\H,\rho,\D)$
and $(\Al,\H,i_u^*\rho,\D + u[\D,u^*] + Ju[\D,u^*]J^\7)$ are
intertwined by the operator $U_u = u J u J^\7$.

This computation entails that the action only depends on the spectral
properties of~$\D$, and strongly suggests to postulate, as Connes has
done recently~\cite{\ConnesGrav}, that the dynamics of real $K$-cycles
is governed by operators of the form $\D + \A + J\A J^\7$, where $\D$
is a ``free'' Dirac operator, and $\A$ denotes a selfadjoint
noncommutative 1-form.

Connes' new dynamical principle seems to run strictly parallel to the
usual gauge principle, but there are important differences. Consider
the commutative $K$-cycle associated to a spin manifold $M$. Then the
only dynamics associated to it in this manner is the trivial one!
First of all, we already remarked that commutative algebras have no
nontrivial inner automorphisms. Secondly, we can check that the
``perturbation'' $\A + J \A J^\7$ vanishes. Indeed, if $\rho(a)$, for
$a \in \Al$, denotes the (left) multiplication operator on the spinor
space, then $J \rho(a^*) J^\7$ is the corresponding right
multiplication operator; these we may identify when $\Al$ is
commutative, writing simply $a = J a^* J^\7$. Now $\A$ is a sum of
terms of the form $a_i\,[\D,b_i]$, and it follows from
$$
\eqalign{
J a\,[\D,b] J^\7 &= a^* J\,[\D,b] J^\7 = J\,[\D,b] J^\7 a^*
\cr
&= [\D,JbJ^\7]\, a^* = [\D,b^*]\, a^* = -\bigl( a \,[\D,b] \bigr)^*
\cr}
$$
that $J \A J^\7 = - \A^* = - \A$. Thus, Connes' principle is a
powerful Ockham's razor: a commutative manifold would support gravity,
but could not support ordinary electromagnetism. In noncommutative
geometry, even {\it abelian\/} gauge fields require the presence of a
noncommutative manifold structure, whose diffeomorphisms incorporate
gauge transformations.

As pointed out also in the previous subsection, commutative algebras
have plenty of {\it outer\/} automorphisms, namely, those given by
diffeomorphisms of the manifold. Invariance under diffeomorphisms is
of course related to gravity ---which in noncommutative spin manifolds
is also naturally described by the neutrino paradigm \cite{\KastlerEH,
\WolfMarkus}.

The strength of Connes' conception is that gauge theories are thereby
deeply connected to the underlying geometry, on the same footing as
gravity. The distinction between gravitational and gauge theories
boils down to the difference between outer and inner automorphisms.
Connes' dynamical principle is not contradicted by experiment, insofar
as it reproduces the Standard Model; and it opens the door to a
formulation in which gravity is intrinsically coupled to the
fundamental interactions of the SM~\cite{\Connesphone, \ChamCon}. In
such a formulation, the Yang--Mills functional is replaced by a
functional interpolating between the model consi\-dered here and the
highly noncommutative geometry at the Planck scale; but that is beyond
the scope of this paper.

Notice that Connes' principle kills the multiplicity of $\theta$~vacua
of~QCD \cite{\Marshak} at the outset.

\subsection{3.5. Minkowskian $K$-cycles}

All treatments of the NCG reconstruction of the Standard Model so far
have been made in the Euclidean framework; at the end one resorts to
the usual incantations to rewrite the orthogonally invariant action
into a Lorentz invariant one. This is not very satisfactory,
especially since the introduction of the new scheme, where charge
conjugation plays a central role.

In order that the real $K$-cycles we use have the good properties, we
employ the physical Dirac operator from the start of calculations.
Actually, the rules of noncommutative differential calculus ---and
therefore the definition of gauge potentials and gauge fields
strength--- need no modification; only the integration rules do, as
$D$ will no longer be elliptic and we are bereft of a Dixmier trace.

For the noncommutative differential forms associated to the product of
an algebra of smooth functions and a matrix algebra, in the compact
euclidean case, the noncommutative integral resolves itself into
finite traces over $\H_F$ and the spinor space, plus the commutative
integral ---where the compactness assumption is removed at the end.

In order to proceed, we need a generalization of \eq{3.3} involving
forms. Take $d = 4$. We have
$$
\(1,f) \propto \int_M \tr f \,\d^4x,
$$
where $f(x)$ is regarded as a scalar matrix in the space of spinors.
In general, for a form $\eta \in \La^k M$ we will have~\cite{\Sirius}:
$$
\(\eta,\eta) \propto \int_M \tr \ga(\eta)^\7 \ga(\eta)
\propto  \int_M \bar\eta \w {\star\eta},
$$
where $\star\eta$ denotes the Hodge dual of $\eta$. For instance, for
a real 2-form $F = \thalf F_{\mu\nu} \,\d x^\mu \w \d x^\nu$ on~$M$,
one obtains $F \w {\star F} = \thalf F_{\mu\nu} F^{\mu\nu} \,\d^4x$.

For forms with values in a matrix algebra, we take traces on the
latter as well, just as in standard nonabelian Yang--Mills theory. For
instance, if
$\eta = \eta^k \ox a_0 + \eta^{k-1} \ox a_1 +\cdots+ \eta^0 \ox a_k$
belongs to $\Om_D^k(\Al)$ for $\Al = C^\infty(M) \ox \Al_F$, then
$$
\(\eta,\eta)  \propto
\int_M  \bar\eta^k \w \star\eta^k \,\tr a_0^\7 a_0
+\cdots+ \bar\eta^0 \w \star\eta^0 \,\tr a_k^\7 a_k.
$$

When $M$ is a Lorentzian manifold, we can define the inner product in
a similar way, using the Hodge dual with the appropriate Lorentzian
signature. Therefore we {\it define\/} the Minkowskian Yang--Mills
action in NCG by $I_{YM} = \(\F,\F)$, where the Lorentz-invariant
scalar product $\(\.,\.)$ on $\Om_D\Al$, with $D$ the {\it physical\/}
Dirac--Yukawa operator, uses the Hodge dual with the Lorentzian
signature, and otherwise is given by the same combination of traces
and commutative integrals as in the Euclidean case. Also, the grading
operator $\ga$ is given by the chirality matrix $\ga_5$ of Dirac
theory. For a Minkowskian spacetime $M$ with a spin structure, we
identify $J$ with the charge conjugation operator $J_C$. Note that on
a four-dimensional space with Lorentzian metric we have
$J_C \ga_5 = - \ga_5 J_C$.

For a more elaborate approach to Minkowskian $K$-cycles,
consult~\cite{\Kalau}.

\subsection{3.6. The $K$-cycle of the Standard Model}

Following Connes and Lott, we propose a particular $K$-cycle able to
produce the whole gamut of the Standard Model particle interactions.
Consider, on the basis of our heuristic considerations in Section~2,
the Eigenschaften algebra
$$
\Al_F := \C \,\oplus \, \HH \,\oplus \, M_3(\C).
$$
Let $\H_F$ have a basis labelled by the list of elementary particles.
The model is given by
$$
\eqalign{
\H &:= L^2(S_M) \ox \H_F \,\simeq\, L^2(\R^4)^4 \ox \H_F,
\cr
\Al &:= C^\infty(M,\R) \ox \Al_F  \simeq  C^\infty(M,\C)
\,\oplus\, C^\infty(M,\HH) \,\oplus\, M_3(C^\infty(M,\C)).
\cr}
$$
The algebra $\Al$ is called the {\it world algebra\/} in this paper.

To obtain the conjugation operator $J$ for this model, we just extend
$J_C$ to~$\H$. The Dirac operator is given by
$$
\D := (\delslash \ox 1) \oplus (1 \ox \D_F),
$$
where we remind the reader that $\ox$ denotes the graded tensor
product. Note, in particular, that this definition guarantees
$\D^2 = (\delslash^2 \ox 1) \,\oplus\, (1 \ox \D_F^2)$. The space
$\Om_\D^1 \Al$ of noncommutative 1-forms is given by
$$
\La^1(M) \ox \Al_F \,\oplus\, C^\infty(M) \ox \Om_{\D_F}^1\Al_F.
$$
The pair $(\Al,\D)$ dictates the matter content of the theory and the
gauge interactions ---and gravity.

The next Section is mainly taken up by the computation of
$\Om_\D^2\Al$, necessary to obtain the action. For these purposes, the
graded tensor product operation will amount to the rule that
$\ga(\d f) \in \La^1(M)$ anticommutes with differentials
$\delta a \in \Om_{\D_F}^1 \Al_F$.

Next we start the step-by-step account of the reconstruction of the
bosonic sector of the SM, with a view to determine which parameter
restrictions, if any, survive the use of the more general scalar
products (see subsection~4.3) allowed by the theory. We have already
specified the finite-part Dirac $\D_F$ ---or Yukawa operator--- in the
space $\H_F$ of the internal degrees of freedom of leptons and quarks.
We start by thoroughly examining the happenings there. This provides
an excellent training ground, and is moreover useful for the complete
model calculation.

\section{4. Reconstructing the bosonic part of the Lagrangian}

\subsection{4.1. The finite-part $K$-cycle}

The representation $\pi$ of $\Al_F = \C \oplus \HH \oplus M_3(\C)$
on~$\H_F$ decomposes into representations on the lepton, quark,
antilepton and antiquark sectors:
$\H_F = \H_F^+ \oplus \H_F^-
= \H_\ell^+ \oplus \H_q^+ \oplus \H_\ell^- \oplus \H_q^-$.

Each of those representations decomposes according to chirality. Thus
$\H_\ell^+$, $\H_q^+$, $\H_\ell^-$ and $\H_q^-$ are graded:
$\H_\ell^+ = \H_{\ell R}^+ \oplus \H_{\ell L}^+$ and so on.

If $q$ is a quaternion, we can write $q = \a + \b j$, where $\a,\b$ are
complex numbers; the conjugate quaternion $q^\7$ is $\bar\a -\b j$. Let
$N$ be the number of fermionic families or generations. For the quark
sector, we have
$$
\H_q^+ = (\C \oplus \C)_R \ox \C^N \ox \C^3_\col
\ \oplus \ \C^2_L \ox \C^N \ox \C^3_\col
$$
and we decide on
$$
\pi_q^+(\la,q,m)
:= \bordermatrix{&\scriptstyle{dR}&\scriptstyle{uR}
&\scriptstyle{dL}&\scriptstyle{uL}\cr
\scriptstyle{dR} & \la &&& \cr
\scriptstyle{uR} && \bar\la && \cr
\scriptstyle{dL} &&& \a & \b \cr
\scriptstyle{uL} &&& -\bar\b & \bar\a \cr} \ox 1_N \ox 1_3
$$
for $\la$ in $\C$, $q$ in $\HH$ and $m$ in $M_3(\C)$. Assuming $N = 3$,
the dimension of $\H_{qR}$ is 18, with basis
$\pmatrix{d_R & u_R & s_R & c_R & b_R & t_R \cr}$, where the colour
indices have been omitted; the dimension of $\H_{qL}$ is also 18, with
standard basis
$\pmatrix {\pmatrix {d'_L & u_L \cr} & \pmatrix {s'_L & c_L \cr}
& \pmatrix {b'_L & t_L \cr} \cr}$. Occasionally we shall denote the
left doublets by~$q_L$.

For the antiquark sector, we have
$$
\H_q^- = (\Bar\C \oplus \Bar\C)_L \ox \Bar\C^N \ox \Bar\C^3_\col
\ \oplus \ \Bar\C^2_R \ox \Bar\C^N \ox \Bar\C^3_\col.
$$
Here we decide on
$$
\pi_q^-(\la,q,m) := 1_4 \ox 1_N \ox m.
$$

For the lepton sector with massless neutrinos, we decide on
$$
\pi_\ell^+(\la,q) :=
\bordermatrix{&\scriptstyle eR &\scriptstyle eL&\scriptstyle\nu L\cr
\scriptstyle eR & \la && \cr
\scriptstyle eL && \a & \b \cr
\scriptstyle\nu L && -\bar\b & \bar\a \cr} \ox 1_N,
$$
on
$$
\H_\ell^+ = \H_{R\ell}^+ \oplus \H_{L\ell}^+
= \C_R \ox \C^N \,\oplus\, \C^2_L \ox \C^N.
$$
The complex dimension of $\H_{\ell R}^+$ is $N$, with standard basis
$\pmatrix{e_R &\mu_R & \tau_R \cr}$; for $\H_{\ell L}^+$ the dimension
is~$2N$, with standard basis
$\pmatrix{\pmatrix{e_L & \nu_e \cr} & \pmatrix{\mu_L & \nu_\mu \cr}
& \pmatrix{\tau_L & \nu_\tau \cr} \cr}$. Occasionally we shall denote
the left doublets by~$\ell_L$.

For the antilepton sector, we take
$\H_\ell^- = \Bar\C_L \ox \Bar\C^N\ \oplus \ \Bar\C^2_R \ox \Bar\C^N$
and $\pi_\ell^-(\la,q,m)$ to be the scalar matrix $\bar\la$.

We summarize the full representation as follows:
$$
\pi(\la,q,m) \pmatrix{\xi \cr \bar\eta \cr}
= \pi(\la,q,m)
\pmatrix{\xi_\ell \cr \xi_q \cr \bar\eta_\ell \cr \bar\eta_q \cr}
= \pmatrix{\pi^+_\ell \xi_\ell \cr \pi^+_q \xi_q \cr
\bar\la \bar\eta_\ell \cr m\bar\eta_q \cr}.
\eqno (4.1)
$$
Within the decomposition of $\H_F^+$ into chiral sectors, $\pi$ yields
a real representation of the complex part of~$\Al_F$, and a real
representation of its quaternionic part. Notice that the
representation of the colour part is non-chiral.

We need to check the real $K$-cycle properties. By construction,
$J_F \pmatrix{\xi \cr \bar\eta \cr} = \pmatrix{\eta \cr \bar\xi \cr}$
so that $J_F^2 = 1$; also, $\D_F J_F = J_F\D_F$ and
$\ga_F J_F = J_F\ga_F$. In order to check that both $\Al_F$ and
$[\D_F, \Al_F]$ commute with $J_F \Al_F J_F^\7$, let us restrict
ourselves to $\H^+_F$ first. Set $a = (\la,q,m)$ and
$b = (\la',q',m')$. It is seen that the action of $J_F b^* J_F^\7$ on
$\H^+_F$ is given by multiplication with the scalar $\la'$ on $\H_\ell$
and by the action of~$m'$ on the colour sector of $\H^+_q$; the
commutation properties \eq{3.5$\,b,c$} follow. To conclude, one can
repeat the arguments on $\H^-_F$ or simply exchange the r\^oles of $a$
and~$b$.

The operator $D_F$ of \eq{3.6} splits into lepton and quark parts:
$$
D_F = \pmatrix{D_\ell & 0 \cr 0 & D_q \ox 1_3 \cr}.
$$
With respect to the right-left splitting, each Dirac operator box is
of the form \eq{3.1}:
$D = \bordermatrix{&\scriptstyle R &\scriptstyle L \cr
\scriptstyle R & 0 & M^\7 \cr \scriptstyle L & M & 0 \cr}$.
Specifically, we choose now
$$
M_\ell = \pmatrix{m_e \cr 0 \cr},  \qquad
M_q = \pmatrix{m_d & 0 \cr 0 & m_u \cr}.
$$
In these formulas $m_d$, $m_u$, $m_e$ are positive definite matrices
of Yukawa coupling constants, acting on the space of generations.
Their eigenvalues are therefore experimentally distinct. With the
basis chosen, one can assume that $m_u$, $m_e$ are diagonalized and
identify: $m_u = m_u^\7$, $m_e = m_e^\7$ if one wishes. On the other
hand, $m_d$ equals its diagonal form premultiplied by the unitary
Kobayashi--Maskawa mixing matrix.

\subsection{4.2. The finite noncommutative differential algebra}

Recall that, to compute the Yang--Mills action, we need the degree-two
part of the noncommutative de~Rham algebra. The representation of the
colour subalgebra commutes with~$\D_F$, and therefore the colour
sector of $\Al_F$ will not contribute to its differential algebra.

Matters are rather different for the flavour subalgebra. When dealing
with finite dimensional $K$-cycles, we shall use $\delta$ to denote
the differentials. As a general matter,
$$
\Al \simeq \Om_D^0\Al \simeq \pi(\Al),  \qquad
\Om_D^1\Al \simeq \pi(\Om^1\Al),  \qquad
\Om_D^2\Al \simeq {\pi(\Om^2\Al) \over \delta(\ker\pi)^1}.
$$

The following computations are done dealing with the quark sector. To
obtain the corresponding formulae for the lepton sector, substitute
$m_d \to m_e$ and $m_u \to 0$. To determine $\Om_{D_F}^1\Al_F$, note
that
$$
\eqalign{
\delta(\la,q,m)
&:= [D_q, \pi_q^+(\la,q,m)]
= \biggl[ \pmatrix{0 & M_q^\7 \cr M_q & 0 \cr},
\pmatrix{\la & 0 \cr 0 & q \cr} \biggr]
\cr
&= \pmatrix{0 & M_q^\7 (q - \la) \cr (\la - q) M_q & 0 \cr}.
\cr}
$$
Here we have used the fact that the complex scalar matrix $\la$
commutes with $M_q$ or $M_q^\7$, whereas the quaternion $q$ and $M_q$
or $M_q^\7$ need not commute. More generally:
$$
(\la_0,q_0,m_0)\, \delta(\la_1,q_1,m_1)
= \pmatrix{0 & M_q^\7 \la_0(q_1-\la_1) \cr q_0(\la_1-q_1) M_q & 0 \cr}.
$$

To find $\Om_{D_F}^2(\Al_F)$, we compute first
$\pi_q(\la_0,q_0,m_0)\,\delta(\la_1,q_1,m_1)\,\delta(\la_2,q_2,m_2)$,
which one sees from the previous equation to be of the form
$$
\pmatrix{M_q^\7 t M_q & 0 \cr 0 & u_1 M_q M_q^\7 u_2 \cr}.
$$
To determine the junk piece $d(\ker\pi)^1$, notice that, for a given
quaternion $q = \a + \b j$,
$$
M_q M_q^\7 q
= \pmatrix{\a & \b \cr -\bar\b & \bar\a \cr}
\ox \thalf(m_dm_d^\7 + m_um_u^\7)
+ \pmatrix{\a & \b \cr \bar\b & -\bar\a \cr}
\ox \thalf(m_dm_d^\7 - m_um_u^\7).
$$
The second summand on the right contains an ``antiquaternion''
$\tilde q := \pmatrix{\a & \b \cr \bar\b & -\bar\a \cr}$; the product
of a quaternion and an antiquaternion is an antiquaternion. The
general element $\sum_j \pi_q^+ (\la_0^j,q_0^j,m_0^j)\,
\delta(\la_1^j,q_1^j,m_1^j)\, \delta(\la_2^j,q_2^j,m_2^j)$ of
$\pi_q^+(\Om^2\Al_F)$ is thus of the form
$$
\pmatrix{M_q^\7 t M_q & 0 \cr
0 & u \ox \thalf(m_dm_d^\7 + m_um_u^\7)
+ \tilde q \ox \thalf(m_dm_d^\7 - m_um_u^\7) \cr}
$$
with $t = \sum_j \la_0^j(q_1^j - \la_1^j)(\la_2^j - q_2^j)$,
$u = \sum_j q_0^j(q_1^j - \la_1^j)(\la_2^j - q_2^j)$, and~$\tilde q$
is an antiquaternion linearly independent of~$t$ and~$u$. This
represents an element in the junk ideal $d(\ker\pi)^1$ if we have
$t = u = 0$: the antiquaternion term need not vanish.

Taking score, we have: $\Al_F = \C \oplus \HH \oplus M_3(\C)$;
$\Om^1_{D_F}\Al_F \simeq \HH \oplus \HH$; and
$\Om^2_{D_F}\Al_F \simeq \HH \oplus \HH$, with the schematic forms:
$$
(r,s) := \pmatrix{& r \cr s & \cr}
:= \pmatrix{0 & M_q^\7 r \cr s M_q & 0 \cr}
\quad \hbox{in } \Om_{D_F}^1 \Al_F,
$$
whereas for $\Om_{D_F}^2\Al_F$, after tossing out the junk we get:
$$
(t,u) := \pmatrix{t & \cr & u \cr}
:= \pmatrix{M_q^\7 t M_q & 0 \cr
0 & u \ox \thalf (m_dm_d^\7 + m_um_u^\7) \cr},
$$
and now the following algebraic rules are transparent:

\item {1.} Involution:
$(\la,q,m)^* = (\bar\la,q^\7,m^\7)$ in $\Al_F$;
$(r,s)^* = (s^\7, r^\7)$ in $\Om^1\Al_F$; $(t,u)^* = (t^\7, u^\7)$
in $\Om^2\Al_F$.
\smallskip

\item {2.} Bimodule structure:
$\Al_F \x \Om_{D_F}^1\Al_F \x \Al_F \to \Om_{D_F}^1\Al_F$:
$$
(\la_1,q_1,m_1) \pmatrix{& r \cr s & \cr} (\la_2,q_2,m_2)
= \pmatrix{& \la_1rq_2 \cr q_1s\la_2 & \cr}
$$
and $\Al_F \x \Om_{D_F}^2\Al_F \x \Al_F \to \Om_{D_F}^2\Al_F$:
$$
(\la_1,q_1,m_1) \pmatrix{t & \cr & u \cr} (\la_2,q_2,m_2)
= \pmatrix{\la_1t\la_2 & \cr &q_1uq_2 \cr}.
$$

Notice the interplay between the involution and the bimodule
operations:
$$
\eqalign{
[(\la_1,q_1,m_1) (r, s) (\la_2,q_2,m_2)]^*
&= (\la_2,q_2,m_2)^* (r, s)^* (\la_1, q_1,m_1)^*,
\cr
[(\la_1,q_1,m_1) (t, u) (\la_2,q_2,m_2)]^*
&= (\la_2,q_2,m_2)^* (t, u)^* (\la_1,q_1,m_1)^*.
\cr}
$$
\smallskip

\item {3.} Multiplication rule:
$\Om_{D_F}^1\Al_F \x \Om_{D_F}^1\Al_F \to \Om_{D_F}^2\Al_F$:
$$
\pmatrix{& r \cr s & \cr} \pmatrix{& t \cr u & \cr}
= \pmatrix{ru & \cr & st \cr}.
$$
\smallskip

\item{4.} The differentials: $\delta\: \Al_F \to \Om_{D_F}^1\Al_F$ and
$\delta\: \Om_{D_F}^1\Al_F \to \Om_{D_F}^2\Al_F$:
$$
\delta(\la,q,m) = (q - \la,\la - q);  \qquad
\delta(r,s) = (r + s, r + s).
$$

Note that $\mp\delta c^* = (\delta c)^*$, for $c \in \Om_{D_F}^k\Al$,
according whether $k$ is even or odd. This way $\pi$ is an involutive
homomorphism.

Recall that $h[(\la_1,q_1,m_1), (\la_2,q_2,m_2)]
:= (\bar\la_1\la_2, q^\7_1 q_2, m^\7_1 m_2)$. By direct computation
one checks the compatibility condition between $\delta$ and $h$:
$$
\delta (\bar\la_1 \la_2, q^\7_1 q_2, m^\7_1 m_2)
= h[(\la_1, q_1, m_1), \delta(\la_2, q_2, m_2)]
- h[\delta(\la_1, q_1, m_1), (\la_2, q_2, m_2)],
$$
in view of the previous remark.  More generally, for a compatible
connection $\nabla$, one must have:
$$
\delta (\bar\la_1 \la_2, q^\7_1 q_2, m^\7_1 m_2)
= h[(\la_1, q_1, m_1), \nabla(\la_2, q_2, m_2)]
- h[\nabla(\la_1, q_1, m_1), (\la_2, q_2, m_2)].
$$

This forces the gauge potentials to be hermitian elements of
$\Om_{D_F}^1\Al_F$, of the generic form:
$$
\A_F = \pmatrix{& M_q^\7 r \cr r^\7 M_q & \cr}.
$$
We pause to indicate that this formula exhibits a ``baby Yukawa''
structure: it is essentially the expression we used at the end of
Section~2, to clarify the r\^ole of the Dirac--Yukawa operator in NCG.

Also, the Yang--Mills action for our toy model exhibits a ``baby
Higgs'' structure. The corresponding gauge field is a hermitian
element of $\Om_{D_F}^2\Al_F$:
$$
\F_F = \delta \A_F + \A^2_F
= (|q_\Phi|^2 - 1) \pmatrix{M_q^\7 M_q & \cr
& \thalf (m_dm_d^\7 + m_um_u^\7) \cr}.
$$
Recall that $1 + r := q_\Phi$. Using the scalar product \eq{3.4} and
adding in the leptonic contribution, the action in terms of the
$q_\Phi$ variable is computed:
$$
\(\F_F,\F_F) = H (|\Phi_1|^2 + |\Phi_2|^2 - 1)^2,
\eqno (4.2)
$$
where
$$
H = {\tfrac92} \tr(|m_d|^4 + |m_u|^4) + 3\tr |m_d|^2|m_u|^2
+ {\tfrac32} \tr |m_e|^4.
$$
One sees that the $SU(2) \x U(1)$ symmetry of the model is broken down
to $U(1)$ by picking the particular minimum of the action.

\subsection{4.3. General scalar products}

The scalar product \eq{3.4} is not unique. This is due to the
existence of superselection sectors. In mathematical terms, the
$K$-cycle $(\Al_F,\H_F,D_F)$ is not irreducible as an
$\Al_F$-bimodule. Let us identify the generic operator $C^+$ on
$\H_F^+$ which commutes with $\pi(\Al_F)$ and
$D_F$. Since $C^+$ commutes with $D_F$, it must be of the form
$C^+ = C_\ell \oplus C_q \ox 1_3$, where $C_j$ commutes with
$\pi_j(\Al_F)$ and $D_j$, for $j = \ell,q$ respectively. The
commutation of $C_\ell$ with $\pi_\ell(\Al_F)$ shows that
$$
C_\ell
= \pmatrix{X_\ell & 0 & 0 \cr 0 & Y_\ell & 0 \cr 0 & 0 & Y_\ell \cr},
$$
where $X_\ell,Y_\ell \in M_N(\C)$. Now $[D_\ell,C_\ell] = 0$ shows
that $m_e X_\ell = Y_\ell m_e$, $m_e^\7 Y_\ell = X_\ell m_e^\7$, and
in particular $[m_e^\7m_e,X_\ell] = 0$. It follows that
$[m_e,X_\ell] = 0$, and hence $Y_\ell = m_e X_\ell m_e^{-1} = X_\ell$;
thus $C_\ell = X_\ell \oplus X_\ell \oplus X_\ell$, and, since
$X_\ell$ is simultaneously diagonalizable with~$m_e$, $C_\ell$ is
determined by the $N$ eigenvalues of~$X_\ell$.

A similar analysis applies to the quark sector; here
$[C_q, \pi_q(\Al_F)] = 0$ implies
$C_q = X_q \oplus X'_q \oplus Y_q \oplus Y_q$, where
$X_q,X'_q,Y_q \in M_N(\C)$.  We see that $[D_q,C_q] = 0$ yields
$m_dX_q = Y_qm_d$, $m_d^\7 Y_q = X_qm_d^\7$, $m_uX'_q = Y_qm_u$ and
$m_u^\7 Y_q = X'_q m_u^\7$. Hence $X_q = Y_q = X'_q$. This matrix
commutes with both $m_d$ and $m_u$; therefore it is a scalar matrix,
on account of flavour mixing.

Consideration of the colour subrepresentation gives no new
restrictions on the particle space. Completely similar arguments on
the antiparticle space $\H^-$ for $C^-$ in the commutant of
$J\pi(\Al_F)J^\7$ gives a $C^- = C'_\ell \oplus C'_q \ox 1_3$ of
exactly the same form. Now, it stands to reason that one should ask
for commutation of $C^+$ with $J\pi(\Al_F)J^\7$ and of $C^-$ with
$\pi(\Al_F)$ and moreover, in view of charge conjugation invariance,
one must have $C^+$ = $C^-$. It is true that, as pointed out
in~\cite{\CIKS}, one obtains a charge conjugation invariant Lagrangian
from the manifestly noninvariant choice $C^+ \neq C^-$ in general; but
this is due to the unphysical doubling of the degrees of
freedom~\cite{\DoubleTrouble} present in all Connes--Lott models till
now; and the latter cannot be considered a permanent feature of the
theory. Our tactics will be to keep $C^+$ and $C^-$ formally separate
till we perform the renormalization analysis in Section~5. The full
matrix $C = C^+ \oplus C^-$, at any rate, belongs then to an algebra
of dimension $2(N + 1)$; the sub-$K$-cycle $(\H_q, D_q \ox 1_3)$ is
irreducible on each flavour or colour sector, whereas
$(\H_\ell, D_\ell)$ has one irreducible component for each lepton
generation. This is related to separate lepton number conservation in
the SM~\cite{\Marshak}. Notice that $C$ will belong to the algebra
generated by $\pi(\Al_F)$ and $J\pi(\Al_F)J^\7$ iff $C_\ell$ and
$C'_\ell$ are also scalars.

Fixing the matrix $C$ is physically equivalent to fixing the coupling
constants in the NCG framework. Note that $C$ is a non-chiral matrix.

Take $C$ to be an arbitrary positive operator belonging to the
commutant of $\pi(\Al_F)$, $J\pi(\Al_F)J^\7$ and $D_F$, as determined
above. More general scalar products agreeable with the theory are
given by
$$
\(\omega,\eta)
:= \Tr(C\omega^\7\eta), \sepword{for} \omega,\eta \in \Om_{D_F}^k\Al,
\eqno (4.3)
$$
and 0~otherwise.  Thus the coefficient $H$ in \eq{4.2} may change.

\medskip

The training course is over. Now for the real thing!

\subsection{4.4. Gauge fields and field strengths for the full algebra}

In order to compute the differential algebra $\Om_D\Al$, we find it
convenient to work with particles and antiparticles separately. We
concentrate on the computation in particle space, the contribution on
the antiparticle subspace being just like in commutative geometry. The
corresponding part of $\Om_D^1\Al$ is generated by elements of the
form $a\,[D,b]$. Writing $a = g_0 \ox (\la_0,q_0)$,
$b = g_1 \ox (\la_1,q_1)$, we get
$$
\eqalign{
a\,[D,b]
&= g_0\ga(\d g_1) \ox (\la_0,q_0)(\la_1,q_1)
+ g_0g_1 \ox (\la_0,q_0) \,\delta(\la_1,q_1) \cr
&= (f_0\ga(\d f_1), r_0\ga(\d r_1))
+ (h_0,s_0)\,\delta(\la_1,q_1),
\cr}
$$
where $f_0,f_1,h_0 \in C^\infty(M,\C)$ and
$r_0,r_1,s_0 \in C^\infty(M,\HH)$; in the notation we have suppressed
terms coming from $C^\infty(M,M_3(\C))$. In matrix notation (for a
schematic splitting of the particle space into chiral sectors) this
can be written:
$$
\pmatrix{f_0 \ga(\d f_1) & M^\7 h_0 (q_1 - \la_1) \cr
s_0 (\la_1 - q_1) M & r_0 \ga(\d r_1) \cr}.
$$

Ignoring also for the time being the contribution of the antiparticle
sector to $\Om_D^2\Al$, we compute the rest of $\pi(\Om^2\Al)$ prior
to junk removal. We notice that
$$
\eqalign{
&\pmatrix{\ga(\d f_1) & M^\7 r_1 \cr s_1 M & \ga(\d q_1) \cr}
\pmatrix{\ga(\d f_2) & M^\7 r_2 \cr s_2 M & \ga(\d q_2) \cr}
\cr
&\qquad = \pmatrix{\ga(\d f_1)\ga(\d f_2) + M^\7 r_1s_2 M
& M^\7 (\ga(\d f_1) r_2 - r_1 \ga(\d q_2)) \cr
(\ga(\d q_1) s_2 - s_1 \ga(\d f_2)) M
& \ga(\d q_1)\ga(\d q_2) + s_1 MM^\7 r_2 \cr},
\cr}
$$
with $f_j \in C^\infty(M,\C)$, $q_j,r_j,s_j \in C^\infty(M,\HH)$; the
minus signs come from the grading of the tensor product (since complex
and quaternion-valued functions need not commute, we will keep the
relative order of the matrix terms instead of interchanging them as in
standard superalgebra calculations).

Junk removal may now proceed efficiently. Computation of
$d(\ker\pi)^1$ shows that the junk ideal consists of terms of the
general form
$$
\pmatrix{\psi \ox 1 & 0 \cr 0 & \chi \ox 1 + \tau \ox (mm^\7)_- \cr}
$$
where $(mm^\7)_-$ is $\thalf(m_dm_d^\7 - m_um_u^\7)$ for the quark
sector, and $\psi$, $\chi$, $\tau$ are functions on~$M$ with complex,
quaternionic and antiquaternionic values, respectively. Thus, two
independent scalar terms drop from $\ga(\d f_1)\ga(\d f_2)$ and
$\ga(\d q_1)\ga(\d q_2)$; the matrix term $MM^\7 r_2$ is converted as
before; and the off-diagonal terms remain unchanged. The product then
becomes
$$
\pmatrix{\ga(\d f_1 \w \d f_2) + (M^\7 r_1s_2 M)_\bot
& M^\7 (\ga(\d f_1) r_2 - r_1 \ga(\d q_2)) \cr
(\ga(\d q_1) s_2 - s_1 \ga(\d f_2)) M
& \ga(\d q_1 \w \d q_2) + (s_1 r_2 \ox (mm^\7)_+)_\bot \cr}
+ \hbox{ junk},
$$
where $(mm^\7)_+$ is $\thalf(m_dm_d^\7 + m_um_u^\7)$ for the quark
sector. The subindex $\perp$ on a matrix indicates that its trace,
which contributes to the junk, has been subtracted out.

Let us now write all terms of the differential. We see that
$d\: \Om_D^0\Al \to \Om_D^1\Al$ is given by:
$$
d(f,q,m) := \pmatrix{\d f & q - f &&\cr f - q & \d q &&\cr
&& \d\bar f &\cr &&& \d m \cr},
$$
where we find it convenient to make explicit the lepton-quark
splitting of the antiparticle space.

The computation of $d\: \Om_D^1\Al \to \Om_D^2\Al$ is slightly more
involved. Indeed, if
$$
\eqalign{
(f_0,q_0,m_0) \,d(f_1,q_1,m_1)
&= \pmatrix{f_0 &&&\cr & q_0 &&\cr && \bar f_0 &\cr &&& m_0 \cr}
\pmatrix{\d f_1 & q_1 - f_1 &&\cr f_1 - q_1 & \d q_1 &&\cr
&& \d\bar f_1 &\cr &&& \d m_1 \cr}
\cr
&= \pmatrix{f_0\,\d f_1 & f_0(q_1 - f_1) &&\cr
q_0(f_1 - q_1) & q_0\,\d q_1 &&\cr
&& \bar f_0\,\d\bar f_1 &\cr &&& m_0\,\d m_1 \cr},
\cr}
$$
then $d(f_0,q_0,m_0) \,d(f_1,q_1,m_1)$ equals
$$
\pmatrix{\d f_0 \w \d f_1 \oplus (q_0 - f_0)(f_1 - q_1)
&\hskip -4pt \d f_0(q_1 - f_1) - (q_0 - f_0)\d q_1 &&\cr
\d q_0(f_1 - q_1) - (f_0 - q_0)\d f_1
&\hskip -4pt \d q_0 \w\d q_1 \oplus (f_0 - q_0)(q_1 - f_1)&&\cr
&&\hskip -6pt \d\bar f_0 \w \d\bar f_1 &\cr
&&&\hskip -6pt \d m_0 \w \d m_1 \cr}
$$
on applying the multiplication rule.

Taking score, we have:
$$
\eqalign{
\Al &\simeq \La^0(M,\C) \oplus \La^0(M,\HH) \oplus \La^0(M,M_3(\C)),
\cr
\Om_D^1\Al &\simeq \La^1(M,\C) \oplus \La^0(M,\HH) \oplus \La^0(M,\HH)
\oplus \La^1(M,\HH) \oplus \La^1(M,M_3(\C)),
\cr
\Om_D^2\Al &\simeq \La^2(M,\C) \oplus \La^0(M,\HH) \oplus \La^1(M,\HH)
\oplus \La^1(M,\HH)
\cr
&\hskip 5.4em      \oplus \La^2(M,\HH) \oplus \La^0(M,\HH)
\oplus \La^2(M,M_3(\C)).
\cr}
$$

The rules of operation are as follows:

1. Involutions:
$$
\eqalign{
\pmatrix{f &&\cr & q &\cr && m \cr}^*
&= \pmatrix{\bar f &&\cr & q^\7 &\cr && m^\7 \cr} \quad\hbox{in $\Al$};
\cr
\pmatrix{A & r &\cr s & V &\cr && M \cr}^*
&= \pmatrix{-\Bar A & s^\7 &\cr r^\7 & -V^\7 &\cr && -M^\7 \cr}
\quad\hbox{in $\Om^1\Al$};
\cr
\pmatrix{K \oplus t & U &\cr Z & L \oplus u &\cr && N \cr}^*
&= \pmatrix{-\Bar K\oplus t^\7 & Z^\7 &\cr U^\7 &-L^\7 \oplus u^\7 &\cr
&& -N^\7 \cr}  \quad\hbox{in $\Om^2\Al$}.
\cr}
$$
The minus signs in these equations are due to our
convention $-dc^* = (dc)^*$, for $c \in \Al$, and the fact that
$dc^* = (dc)^*$, for $c \in \Om_D^1\Al$.

2. Bimodule operations:
$\Al \x \Om_D^1\Al \x \Al \to \Om_D^1\Al$:
$$
\pmatrix{f_1 &&\cr & q_1 &\cr && m_1 \cr}
\pmatrix{A & r &\cr s & V &\cr && M \cr}
\pmatrix{f_2 &&\cr & q_2 &\cr && m_2 \cr}
= \pmatrix{f_1 A f_2 & f_1 r q_2 &\cr q_1 s f_2 & q_1 V q_2 &\cr
&& m_1 M m_2 \cr}
$$
and $\Al \x \Om_D^2\Al \x \Al \to \Om_D^2\Al$:
$$
\eqalign{
\pmatrix{f_1 &&\cr & q_1 &\cr && m_1 \cr}
&\pmatrix{K \oplus t & U &\cr Z & L \oplus u &\cr && N \cr}
\pmatrix{f_2 &&\cr & q_2 &\cr && m_2 \cr}
\cr
= &\pmatrix{f_1 K f_2 \oplus f_1 t f_2 & f_1 U q_2 &\cr
q_1 Z f_2 & q_1 L q_2 \oplus q_1 u q_2 &\cr && m_1 N m_2\cr}
\cr}
$$

3. Multiplication rule: the multiplication rule for
$\Om_D^1\Al \x \Om_D^1\Al \to \Om_D^2\Al$ is written:
$$
\pmatrix{A_1 & r_1 &\cr s_1 & V_1 &\cr && M_1 \cr}
\pmatrix{A_2 & r_2 &\cr s_2 & V_2 &\cr && M_2 \cr}
= \pmatrix{A_1 \w A_2 \oplus r_1 s_2 & A_1 r_2 - r_1 V_2 &\cr
V_1 s_2 - s_1 A_2 & V_1 \w V_2 \oplus s_1 r_2 &\cr
&& M_1 M_2 \cr}.
$$

4. The differentials: $d\: \Al \to \Om_D^1\Al$ and
$d\: \Om_D^1\Al \to \Om_D^2\Al$ are given by:
$$
\eqalign{
d\pmatrix{f &&\cr & q &\cr && m \cr}
&:= \pmatrix{\d f & q - f &\cr f - q & \d q &\cr && \d m \cr},
\cr
d\pmatrix{A & r &\cr s & V &\cr && M \cr}
&:= \pmatrix{\d A \oplus (r + s) & \d r + A - V &\cr
\d s - A + V & \d V \oplus (r + s) &\cr && \d M \cr}.
\cr}
$$

The smoke has cleared and one can appreciate the beautifully simple
structure of Connes' algebra. Here is perhaps the point to mention the
superconnection approach to the Standard Model, evolved by the
Marseille--Mainz group~\cite{\Coque, \Mainz}. Their {\it d\'emarche\/}
was in some sense the reverse of what is done here, namely, they
started from the bosonic sector of the SM Lagrangian, from which they
read an algebraic structure similar to the one derived here. In their
approach, information about the Dirac--Yukawa operator is inferred
from the bosonic sector.

Noncommutative gauge potentials are selfadjoint elements of
$\Om_D^1\Al$, hence of the form
$$
\A = \pmatrix{B & q_\Phi - 1 &\cr q_\Phi^\7 - 1 & W &\cr && A \cr}.
$$

\subsection{4.5. Gauge reduction rule and assignment of hypercharges}

By construction, the theory up to now is invariant under the unitary
group $\U(\Al) = C^\infty(M, U(1) \x SU(2) \x U(3))$. The unitary
group of the quaternions is $\U(\HH) = SU(2)$, of course. We must find
a way to reduce this group to the SM group
$C^\infty(M, SU(2) \x U(1)_Y \x SU(3))$, where $Y$ stands for weak
hypercharge.

The matter is dealt with by imposing the so-called {\it unimodularity
condition\/} on the noncommutative gauge potential:
$$
\Tr(\A + J\A J^\7)^+ = 0,
$$
(the superscript $+$ denoting restriction to particle space), that
implies $\Tr(\A + J\A J^\7)^- = 0$. We shall explain how it works
before discussing it any further. The diagonal part of
$(\A + J\A J^\7)^+$ is of the form
$$
\bordermatrix{&\scriptstyle{eR}&\scriptstyle{dR}&\scriptstyle{uR}
&\scriptstyle{\ell L}&\scriptstyle{qL}\cr
\scriptstyle{eR}     & B + B & & & & \cr
\scriptstyle{dR}     & & B + A & & & \cr
\scriptstyle{uR}     & & & -B + A & & \cr
\scriptstyle{\ell L} & & & & W + B & \cr
\scriptstyle{qL}     & & & & & W + A\cr}
\eqno (4.4)
$$
where $B$, $W$, $A$ are the respectively $U(1)$, $SU(2)$ and
$U(3)$-valued gauge potentials. Note that $W$ is traceless. Therefore,
the unimodularity condition reads:
$$
B + \tr A = 0.
$$
With that condition, the abelian part of the noncommutative gauge
potential reads:
$$
\bordermatrix{&\scriptstyle{eR}&\scriptstyle{dR}&\scriptstyle{uR}
&\scriptstyle{\ell L}&\scriptstyle{qL}\cr
\scriptstyle{eR}     & 2B & & & & \cr
\scriptstyle{dR}     & & \tfrac23 B & & & \cr
\scriptstyle{uR}     & & & -\tfrac43 B & & \cr
\scriptstyle{\ell L} & & & & B & \cr
\scriptstyle{qL}     & & & & & -\tfrac13 B\cr}
\oplus\,
\bordermatrix{&\scriptstyle{\bar eL}&\scriptstyle{\bar dL}
&\scriptstyle{\bar uL}&\scriptstyle{\bar\ell R}
&\scriptstyle{\bar qR}\cr
\scriptstyle{\bar eL}    & -2B & & & & \cr
\scriptstyle{\bar dL}    & & -\tfrac23 B & & & \cr
\scriptstyle{\bar uL}    & & & \tfrac43 B & & \cr
\scriptstyle{\bar\ell R} & & & & -B & \cr
\scriptstyle{\bar qR}    & & & & & \tfrac13 B\cr}.
$$
We get, into the bargain, the tableau of hypercharge assignments of
the SM: if we conventionally adopt $Y(e_L,\nu_L) = -1$, there follows
for left-handed (anti-)leptons and \hbox{(anti-)}\allowbreak
quarks: $Y(\bar e_L) = 2$; $Y(\bar d_L) = \tfrac23$;
$Y(\bar u_L) = -\tfrac43$ and $Y(d_L, u_L) = \tfrac13$.

\smallskip

Therefore, the unimodularity prescription works fine. However, it
always struck some people, including Connes, as an artificial
appendage. The new dynamical principle of Connes, by postulating that
the only physically admissible gauge fields are the ones that appear
as (inner) automorphisms of the noncommutative manifold itself, gives
a rationale for unimodularity: the trace of $\A + J\A J^\7$
contributes only a scalar term to the $K$-cycle, that according to
\eq{3.2} does not modify the metric (this is related to the global
$U(1)$-invariance of the Lagrangian).

As pointed out in the introduction, an apparently different tack was
taken in \cite{\Chiron}, where the unimodularity condition was found
equivalent, within the present NCG model for the SM, to cancellation
of anomalies. A first exciting hint at a deeper relationship between
quantum physics and NCG? We return to the matter in Section~6.

\subsection{4.6. The noncommutative Yang--Mills action}

The noncommutative gauge field is written:
$$
\F =: \pmatrix{F_{\ell f} &&& \cr & F_{qf} && \cr
&& F_{\ell c} & \cr &&& F_{qc}},
\eqno (4.5)
$$
with a lepton--quark splitting: a matrix of (ordinary) differential
2-, 1- and 0-forms.  In \eq{4.5} we have
$$
\eqalign{
F_{qf}
&= \pmatrix{\d B \oplus (qq^\7 - 1) & \d q + Bq - qW \cr
\d q^\7 - q^\7 B + Wq^\7 & (\d W + W\w W) \oplus (q^\7q - 1)\cr}
\ox 1_N \ox 1_3,
\cr
F_{qc} &= \pmatrix{1_2 & 0 \cr 0 & 1_2\cr} \ox 1_N \ox (\d A + A\w A),
\cr
F_{\ell f}
&= \pmatrix{\d B \oplus (qq^\7 - 1) & \d q + Bq - qW \cr
\d q^\7 - q^\7 B + Wq^\7 & (\d W + W\w W) \oplus (q^\7q - 1)\cr}
\ox 1_N,
\cr
F_{\ell c} &= \pmatrix{-\d B & 0 \cr 0 & -\d B \ox 1_2 \cr} \ox 1_N,
\cr}
$$
where a $3 \x 3$ matrix notation is understood in the electroweak
sector for the last two formulas.

In order to compute the Lagrangian, one has to compute the different
terms of \hbox{$\(\F,\F)$}. Let us forget about leptons temporarily.
Rename $C_q$ to $C_{qf}$ and $C'_q$ to $C_{qc}$, as the respective
coefficients of the flavour and the colour gauge fields for the quark
sector. Also, write simply $F_{qf,ij}^2$ for the coefficient of
$\d^4x$ in $F_{qf,ij} \w \star F_{qf,ij}$ ---a nonpositive quantity in
general, on account of the use of the Minkowskian Hodge product--- and
similarly for $F_{qc}$.

Then the contribution of the quark sector to the bosonic part of the
action is given by
$$
I_q = \int_M (\L_{2qf} + \L_{2qc} + \L_{1q} + \L_{0q}) \,\d^4x,
$$
where $\L_{2qf} + \L_{1q} + \L_{0q} = C_{qf} \sum_{ij} F_{qf,ij}^2$
and $\L_{2qc} = C_{qc} \sum_{ij} F_{qc,ij}^2\,$: we have denoted by
$\L_{2qf}$, $\L_{1q}$, $\L_{0q}$ respectively, the terms that come
from 2-forms, 1-forms and 0-forms in the NCG flavour field; and by
$\L_{2qc}$ the term coming from the NCG colour field.

The chromodynamics part $\L_{2qc}$ is worked out like in commutative
geometry. We can write $A = A^a \,\lab_a/2$, where $\row \lab18$ are
always the Gell-Mann matrices, and $\lab_0 = (2/\sqrt3)1_3$. (Recall
that $\tr(\lab_a\lab_b) = 2 \delta_{ab}$ and
$[\lab_a,\lab_b] = 2 f_{ab}{}^c \,\lab_c$.) We have then
$\d A + A\w A = (\d A^c + \thalf f_{ab}{}^c\, A^a \w A^b)\,\lab_c/2$.
In view of orthogonality, the nine components of $\d A + A \w A$ make
independent contributions to the Lagrangian $\L_{2q}$. Let us write,
for $c = 1,2,\dots,8$:
$$
A^c = i g_3 A_\mu^c \,dx^\mu, \qquad
G_{\mu\nu}^c := \del_\mu A_\nu^c - \del_\nu A_\mu^c
+ i g_3\, f_{ab}{}^c A_\mu^a A_\nu^b.
$$
These terms contribute the gluon field:
$$
\L_{2qc} := 4 C_{qc} N \sum_{c=1}^8
((\d A^c + \thalf f_{ab}{}^c \,A^a \w A^b) \,\lab_c/2)^2
= -4 C_{qc} N (\tquarter g_3^2 G_{\mu\nu}^c G^{\mu\nu}_c).
$$

The matrix whose traced square gives the flavour contribution to
$\L_{2q}$ is
$$
\pmatrix{\d B & 0 && \cr 0 & -\d B && \cr
&& \d W_1 - W_2 \w W_2^* & \d W_2 + 2 W_1 \w W_2 \cr
&&-\d W_2^* - 2 W_2^* \w W_1 & -\d W_1 + W_2 \w W_2^*\cr}
\ox 1_N \ox 1_3.
$$
That yields
$$
\L_{2qf} = 6 C_{qf} N \bigl((\d B)^2 + (\d W_1 - W_2 \w W_2^*)^2
+ (\d W_2 + 2 W_1 \w W_2)^2 \bigr).
$$

Set, as is the custom,
$$
F_{\mu\nu}^a
:= \del_\mu W_\nu^a - \del_\nu W_\mu^a
+ g_2\eps^a{}_{bc} W_\mu^b W_\nu^c,  \qquad
B_{\mu\nu} := \del_\mu B_\nu - \del_\nu B_\mu.
$$
With this in hand, we compute
$$
\d W_2 + 2 W_1 \w W_2
= -\tiquarter g_2 (F_{\mu\nu}^1 - i F_{\mu\nu}^2) \,dx^\mu \w dx^\nu.
$$
Next,
$$
\d W_1 - W_2 \w W_2^*
= -\tiquarter g_2 F_{\mu\nu}^3 \,dx^\mu \w dx^\nu.
$$
Also, $\d B = \tiquarter g_1 B_{\mu\nu} \,dx^\mu \w dx^\nu$.

We square and add the several 2-forms, arriving finally at
$$
\L_{2qf} = 3 C_{qf} N \bigl( \tquarter g_1^2 B_{\mu\nu} B^{\mu\nu}
+ \tquarter g_2^2 F_{\mu\nu}^a F_a^{\mu\nu} \bigr).
$$

The ``diagonal'' colour term
$(\d A + A \w A)^0 = \d A^0 \ox 1_3/\sqrt3$ must be added now. Notice
that $A_{11} + A_{22} + A_{33} = \sqrt3\,A^0$ (since the $\lab_a$ are
traceless); the reduction condition for the quark sector is then
$B = -\sqrt3\,A^0$. Thus we must add the term $A^0/\sqrt3 = -\tthird B$
to each diagonal entry, which gives rise to a term of the form
$\tfrac16 C_{qc} N g_{1}^2 B_{\mu\nu} B^{\mu\nu}$.

Let us abbreviate $t_q := \tr(m_dm_d^\7 + m_um_u^\7)$. The quark
sector contribution to the kinetic term for the Higgs fields is:
$$
\eqalign{
\L_{1q} &= 2 C_{qf} t_q (\d q + Bq - qW)^2
\cr
&= 6 C_{qf} t_q \bigl((\d\Phi_1 + (W_1 - B)\Phi_1 + W_2\Phi_2)^2
+ (\d\Phi_2 - (W_1 + B)\Phi_2 - W_2^*\Phi_1)^2 \bigr)
\cr
&= 6 C_{qf} t_q ({\bm D}\Phi)^2 = 6 C_{qf} t_q\, D_\mu\Phi\, D^\mu\Phi,
\cr}
$$
where ${\bm D} = \d - B + W$ is the covariant derivative for the Higgs
field. With the identifications previously made, we have $D_\mu
= \del_\mu - \thalf ig_1 B_\mu - \thalf ig_2\,\taub\cdot{\bm W}_\mu$.
This is consistent with the hypercharge $+1$ of the Higgs, already
determined.

The last curvature terms yield the Higgs self{}interaction term:
$\L_{0q} = C_{qf} H_q (\|\Phi_1\|^2 + \|\Phi_2\|^2 - 1)^2$, where
$$
H_q = 3 \tr \pmatrix{((m_d^\7m_d)_\bot)^2 \cr
&\!\!\! ((m_u^\7m_u)_\bot)^2 \cr
&&\!\!\! \tquarter((m_dm_d^\7 + m_um_u^\7)_\bot)^2 \cr
&&&\!\!\! \tquarter((m_dm_d^\7 + m_um_u^\7)_\bot)^2 \cr}
$$
By taking separately the traces on $\H_R$ and $\H_L$ to determine the
scalar junk terms, one sees that
$(m_d^\7m_d)_\bot = m_d^\7m_d - t_q/2N$,
$(m_u^\7m_u)_\bot = m_u^\7m_u - t_q/2N$, and
$(m_dm_d^\7 + m_um_u^\7)_\bot = m_dm_d^\7 + m_um_u^\7 - t_q/N$. Then
we find that
$$
\L_{0q}
= \bigl( \tfrac92 \tr(m_dm_d^\7)^2 + \tfrac92 \tr(m_um_u^\7)^2
+ 3\tr(m_dm_d^\7m_um_u^\7) - 3t_q^2/N \bigr)
\bigl( \|\Phi_1\|^2 + \|\Phi_2\|^2 - 1 \bigr)^2.
$$

Note that $H_q = 0$ when $N = 1\,$: in the NCG treatment, a
symmetry-breaking sector can exist only if the number of fermion
families is greater than one.

To finish the computation, assume now there are only leptons in the
world. As advertised, the terms in the Lagrangian corresponding to $N$
massless neutrino generations have different coefficients for each
generation. Let us put $C_{\ell f} := \tr C_\ell/N$ and
$C_{\ell c} := \tr C'_\ell/N$.  We get for the flavour part:
$$
\L_{2\ell f}
= \thalf C_{\ell f} N \bigl( \tquarter g_1^2 B_{\mu\nu} B^{\mu\nu}
+ \tquarter g_2^2 F_{\mu\nu}^a F_a^{\mu\nu} \bigr).
$$
The lepton sector contribution to the kinetic term for the Higgs
fields is:
$$
\L_{1\ell} = 2 C_{\ell f} t_\ell \,D_\mu\Phi \,D^\mu\Phi,
$$
where $\ell$ stands for the said leptons, $t_\ell := \tr(m_em_e^\7)$
and the lepton sector contribution to the Higgs self{}interaction term
is
$$
\L_{0\ell} = (\|\Phi_1\|^2 + \|\Phi_2\|^2 - 1)^2
\, \bigl( \tfrac32 (m_em_e^\7)^2 - t_\ell^2/N \bigr).
$$
The abelian leptonic part of the colour field contributes a term of
the form
$$
\L_{2\ell c}
= \tfrac32 C_{\ell c} N (\tquarter g_1^2 B_{\mu\nu} B^{\mu\nu}).
$$

Now we have to combine the lepton and quark sectors. One ought to be
careful here, because this combination is not entirely
straightforward; our following formulae correct an oversight in a
previous publication~\cite{\Sirius}. We would expect the general
action functional to be a combination of the contributions of the
quark and lepton sectors, with different coefficients for the $F_f$
and $F_c$ sectors. In the general Lagrangian $\L = \L_0 + \L_1 + \L_2$
the gauge and kinetic terms combine linearly. On the other hand,
{\it the Higgs self{}interaction terms do not combine linearly}, on
account of the junk removal, and the question of the precise form of
the $\L_0$ term is tangled with the choice of the scalar product.
Write $F_{0f} = F_{0R} \oplus F_{0L}$ for the part of the NCG flavour
gauge field which contributes to the Higgs self{}interaction term,
decomposed according to its chiral components. If we abbreviate
${\bm H} := \|\Phi_1\|^2 + \|\Phi_2\|^2 - 1$, we have:
$$
\eqalign{
F_{0R} &= {\bm H} \biggl[ (m_e^\7m_e - \psi \ox 1_N) \oplus
\pmatrix{m_d^\7m_d - \psi \ox 1_N & 0 \cr
0 & m_u^\7m_u - \psi \ox 1_N \cr} \ox 1_3 \biggr],
\cr
F_{0L} &= {\bm H} \biggl[
\pmatrix{\thalf m_em_e^\7 - \chi \ox 1_N & 0 \cr
0 & \thalf m_em_e^\7 - \chi \ox 1_N \cr}
\cr
&\qquad\qquad \oplus
\pmatrix{\thalf(m_dm_d^\7 + m_um_u^\7) - \chi \ox 1_N & 0 \cr
0 & \thalf(m_dm_d^\7 + m_um_u^\7) - \chi \ox 1_N \cr}\ox 1_3
\biggr],
\cr}
$$
after subtraction of the junk terms $\psi$, $\chi$. On account of the
form \eq{4.1} of the representation of the algebra, we must also
subtract corresponding junk terms from the lepton-colour sector:
$$
F_{0c} = {\bm H} \pmatrix{ - \bar\psi \ox 1_N & 0 \cr
0 & - \bar\psi \ox 1_N \ox 1_2 \cr}.
$$
The junk terms are determined by the requirement that the residual
terms be orthogonal to the junk ideal, which here simply means that
the above matrices are traceless for the operative trace
$\tr_\ell C_\ell + C_{qf} \tr_q + \tr_\ell C'_\ell + C_{qc} \tr_q$.

Then we prescribe:
$$
\eqalign{
0 &= (\tr(C_\ell m_e^\7m_e) - C_{\ell f} N\psi)
+ 3C_{qf}(\tr(m_d^\7m_d) + \tr(m_u^\7m_u) - 2N\psi)
- 3C_{\ell c} N\bar\psi
\cr
&= - (C_{\ell f} + 6 C_{qf}) N\psi - 3C_{\ell c} N\bar\psi
+ (\tr(C_\ell m_e^\7m_e) + 3C_{qf} \tr(m_d^\7m_d + m_u^\7m_u)),
\cr}
$$
which happens iff $\psi$ is real and equals
$$
\psi = {1\over C_{\ell f} + 6 C_{qf} + 3 C_{\ell c}} {t \over N},
\sepword{with}
t := \tr(C_\ell m_e^\7m_e) + 3C_{qf} \tr(m_d^\7m_d + m_u^\7m_u);
$$
and
$$
\eqalign{
0 &= 2 (\tr(\thalf C_\ell m_em_e^\7) - N\chi)
+ 6 C_{qf} (\tr(\thalf(m_dm_d^\7 + m_um_u^\7)) - N\chi)
\cr
&= - (2 C_{\ell f} + 6C_{qf}) N\chi + t,
\cr}
$$
which happens iff
$$
\chi = {1 \over 2 C_{\ell f} + 6 C_{qf}} {t \over N}.
$$

On computing $\ncint F_0^2$, we now obtain:
$$
\eqalign{
{\bm H}^{-2} \L_0
&= \tr\bigl( (C_\ell m_e^\7m_e - \psi\ox 1)^2
+ 2(\thalf m_em_e^\7 - \chi\ox 1)^2 \bigr)
+ 3C_{qf} \tr\bigl((m_d^\7m_d - \psi\ox 1)^2
\cr
&\qquad + (m_u^\7m_u - \psi\ox 1)^2
+ 2(\thalf(m_dm_d^\7 + m_um_u^\7) - \chi\ox 1)^2
+ 3C_{\ell c} (\psi\ox 1)^2 \bigr)
\cr
&= \tfrac32 \tr(C_\ell m_e^\7m_e)^2
+ \tfrac92 C_{qf} \tr((m_d^\7m_d)^2 + (m_u^\7m_u)^2)
+ 3C_{qf} \tr(m_d^\7m_dm_u^\7m_u)
\cr
&\qquad - 2t(\psi + \chi) + N\psi^2(C_{\ell f} + 6C_{qf} + 3C_{\ell c})
+ N\chi^2 (2C_{\ell f} + 6C_{qf})
\cr
&= \tfrac32 \tr(C_\ell m_e^\7m_e)^2 + \tfrac92 C_{qf}\tr((m_d^\7m_d)^2
+ (m_u^\7m_u)^2) + 3C_{qf} \tr(m_d^\7m_dm_u^\7m_u)
\cr
&\qquad- {t^2\over N} \Bigl({1\over C_{\ell f} + 6C_{qf} + 3C_{\ell c}}
+ {1\over 2C_{\ell f} + 6C_{qf}} \Bigr).
\cr}
$$

Putting it all together, we arrive at the bosonic part of the
Connes--Lott Lagrangian.  One obtains:
$$
\eqalignno{
\L &= - \tfrac14 A B_{\mu\nu} B^{\mu\nu}
- \tfrac14 E F_{\mu\nu}^a F_a^{\mu\nu}
- \tfrac14 C G_{\mu\nu}^a G_a^{\mu\nu} + S D_\mu\Phi\, D^\mu\Phi
\cr
&\qquad - L (\|\Phi_1\|^2 + \|\Phi_2\|^2)^2
+ 2L (\|\Phi_1\|^2 + \|\Phi_2\|^2),
& (4.6) \cr}
$$
where
$$
\eqalign{
A &= N (\thalf C_{\ell f} + \tfrac32 C_{\ell c} + 3 C_{qf}
+ \tfrac23 C_{qc}) g_1^2,
\cr
E &= N (C_{\ell f} + 3 C_{qf}) g_2^2,
\cr
C &= 4 N C_{qc} g_3^2,
\cr
S &= 2 \tr(C_\ell m_e^\7m_e) + 6C_{qf} \tr(m_d^\7m_d + m_u^\7m_u),
\cr
L &= \tfrac32 \tr(C_\ell m_e^\7m_e)^2
+ \tfrac92 C_{qf} \tr((m_d^\7m_d)^2 + (m_u^\7m_u)^2)
+ 3C_{qf} \tr(m_d^\7m_d m_u^\7m_u)
\cr
&\enspace - {1\over N}\Bigl({1\over C_{\ell f} + 6C_{qf} + 3C_{\ell c}}
+ {1\over 2C_{\ell f} + 6C_{qf}}\Bigr)
(\tr(C_\ell m_e^\7m_e) + 3C_{qf} \tr(m_d^\7m_d + m_u^\7m_u))^2.
\cr}
$$

\section{5. The NCG constraints: a first confrontation with experiment}

\subsection{5.1. The constraints at the tree level}

It remains to identify the parametrized SM Lagrangian~\eq{4.6} to the
usual SM Lagrangian
$$
\eqalign{
\L &= - \tquarter B_{\mu\nu} B^{\mu\nu}
- \tquarter F_{\mu\nu}^a F_a^{\mu\nu}
- \tquarter G_{\mu\nu}^a G_a^{\mu\nu} + D_\mu\phi\, D^\mu\phi
\cr
&\qquad - \la (\|\phi_1\|^2 + \|\phi_2\|^2)^2
+ m^2 (\|\phi_1\|^2 + \|\phi_2\|^2),
\cr}
$$
Recall that $\Phi = \sqrt2\,\phi/v$.  Note that the equality
$\la v^2 = m^2$ is an automatic consequence of this identification.

To continue, we stipulate that the term
$S D_\mu\Phi\, D^\mu\Phi = (2S/v^2) D_\mu\phi\, D^\mu\phi$ be none
other than the usual Higgs kinetic term $D_\mu\phi\, D^\mu\phi$. Thus
we set $S = \thalf v^2$. At this point, we make the very good
approximation of neglecting all fermion masses except for the top mass
$\mtop$, whereupon $S \simeq 6 C_{qf}\,\mtop^2$. (For the ``fuzzy
relations'' that one obtains when this approximation is not made,
consult~\cite{\Fuzzy}.)

Setting $E = 1$ (or equivalently, $\thalf v^2 E = S$) gives a relation
for the mass of the intermediate vector boson $m_W = \thalf g_2 v$:
$$
m_W = \mtop \sqrt{{1\over N}\,{3C_{qf} \over C_{\ell f} + 3C_{qf}}}.
\eqno (5.1)
$$
{}From this one gets the constraint $\mtop \geq \sqrt N\,m_W$. With
$N = 3$ and $m_W = 80.2\GeV$, that would give $\mtop \gtrsim 139\GeV$,
right in the ballpark. Notice also that with the present socially
accepted value of $\mtop = 175\GeV$~\cite{\Particledata},
one would obtain a theoretical constraint%
\footnote*{We owe this remark to M. Quir\'os.}
on the number of generations: $N \lesssim 5$.

Kastler and Sch\"ucker introduced~\cite{\ArtstateIV} the concept of
``lepton dominance'' to characterize the situation in which
$C_{\ell f} > C_{qf}$. Extreme ``lepton dominance'' in the flavour
sector is tantamount to huge top masses. In order to quantify it, they
introduce the parameter
$x := (C_{\ell f} - C_{qf})/(C_{\ell f} + C_{qf})$, with range
$-1 \leq x \leq 1$, and lepton dominance corresponds to $x > 0$. The
formula \eq{5.1} becomes
$$
\mtop = m_W \sqrt{{N \over 3}\, {4 - 2x \over 1 - x}}.
\eqno (5.2)
$$
Note that lepton dominance, with $N = 3$ generations, would correspond
to $\mtop > 2m_W \approx 160.4\GeV$.

Similarly, the relation $C = 1 = E$ yields
$$
g_3 = \thalf g_2 \sqrt{{C_{\ell f} + 3C_{qf} \over C_{qc}}}.
\eqno (5.3)
$$

For the Weinberg angle $\theta_W$, we may express
$\sin^2\theta_W = g_1^2/(g_2^2 + g_1^2)$ from the relation $A = 1 = E$
and we get:
$$
\sin^2 \theta_W
= {C_{\ell f} + 3C_{qf} \over \tfrac32 C_{\ell f} + \tfrac32 C_{\ell c}
+ 6 C_{qf} + \tfrac23 C_{qc}}.
\eqno (5.4)
$$

Finally, the mass of the Higgs $m_H = \sqrt2\,\mu$ obeys
$m_H^2 = 4L/S$, where $L \simeq \tfrac92 C_{qf} \mtop^4
- (9C_{qf}^2 \mtop^4/N) \bigl((C_{\ell f} + 6C_{qf} + 3C_{\ell c})^{-1}
+ (2C_{\ell f} + 6C_{qf})^{-1}\bigr)$. This yields
$$
m_H = \mtop \sqrt{3 - {6\over N}
\Bigl( {C_{qf}\over C_{\ell f} + 6C_{qf} + 3C_{\ell c}}
+ {C_{qf}\over 2C_{\ell f} + 6C_{qf}} \Bigr)}.
$$

Several preliminary conclusions are warranted by the foregoing. Note
that, as long as we do not set $C^+ = C^-$, and discounting the
overall multiplicative constant, the NCG approach has in practice
three unknown constants to fit four relations between the parameters
of the SM. In the original analysis \cite{\ArtstateIV}, where (in our
notation) $C_{\ell f} = C_{\ell c}$ and $C_{qf} = C_{qc}$, a clash
between \eq{5.3} and \eq{5.4} {\it prima facie\/} and experiment was
thus softened only in the extreme lepton dominance limit ---then huge
top masses ensued. Eq.~\eq{5.3} does not constrain, however, the
values of $g_3$ and $g_2$ to a given range, since $C_{qc}$ is an
arbitrary positive number. Finally, since $C_{\ell c}$ is also
arbitrary, \eq{5.4} only sets an upper bound for $\sin^2\theta_W$
of~$2/3$. In other words, the unphysical distinction between the
noncommutative coupling constants in the particle and the antiparticle
sectors served to accommodate the experimental values of the strong
coupling constant and the Weinberg angle by conveniently shifting part
of the ``lepton dominance'' to the colour sector. Nowadays, the need
for renormalization of constraints coming from noncommutative geometry
is generally recognized. Thus, there is no cogent reason to assume
$C_{\ell f} \neq C_{\ell c}$ and $C_{qf} \neq C_{qc}$. In what
follows, we drop the distinction between $C^+$ and $C^-$. Then,
putting in the previous equation $C_{\ell c} = C_{\ell f}$ and taking
into account the definition of $x$, we end up with the formula
$$
\mtop \sqrt{3 - {9\over 2N}
\Bigl( {x^2 - 4x + 3 \over x^2 - 7x + 10} \Bigr)}
\eqno (5.5)
$$
for the Higgs particle mass. From this, with $N = 3$, we would get the
relatively tight constraints
$$
\sqrt{7/3}\,\mtop \leq m_H \leq \sqrt{3}\,\mtop.
$$

In summary, in the absence of a good criterion to fix the more
important parameter of the theory, namely~$x$, the {\it classical\/}
Lagrangian of the Standard Model has, within the noncommutative
geometry formalism, seventeen free parameters rather the conventional
eighteen. On the other hand, the experimentally determined top mass
helps to determine the possible values of~$x$. Moreover, once the top
mass is fairly well pinned down, the model seems to fix the value of
the Higgs mass. For instance, with $m_W \simeq 80.3\GeV$, then if
$\mtop = 2\,m_W \simeq 160.6\GeV$, from \eq{5.2} we get $x = 0$, and
then from \eq{5.5} that $m_H \simeq 256.5\GeV$. If
$\mtop = 2.5\,m_W \simeq 200.9\GeV$, we get $x \simeq 0.53$, and then
it would follow that $m_H \simeq 332.7\GeV$. If
$\mtop = 2.25\,m_W \simeq 181\GeV$, we get $x \simeq 0.35$, and in
that case $m_H \simeq 295.4\GeV$. For the value
$\mtop = (175 \pm 12)\GeV$ accepted at present~\cite{\Particledata},
$m_H$ should roughly lie in the interval $250{-}324\GeV$.

At any rate, noncommutative geometry clearly points to the Higgs mass
being of the same order of magnitude as, but somewhat bigger than, the
highest fermion mass. Note that for $x$ greater than about $0.9$,
corresponding in the model to a top mass of about $375\GeV$, we would
be outside the perturbative regime in quantum field theory. If there
were a compelling reason to adopt Connes' relations on-shell, the
model considered in this paper would stand or fall by the value of the
Higgs mass.

\subsection{5.2. Renormalization analysis}

The problem we face next is the formulation of a quantum theory whose
classical action is given by \eq{2.3} plus the NCG Yang--Mills action
$I_{YM} = \(\F,\F)$. Unfortunately, no quantization procedure has been
developed yet within the framework of NCG. We can, of course, take the
classical Lagrangian given by \eq{4.6}, and use conventional
quantization methods to obtain a quantum field theory. This is not, in
principle, a satisfactory state of affairs, for it would appear that
the standard quantization methods keep no record of the fact that
gauge and Higgs fields constitute a single noncommutative gauge field,
their quantization being carried out independently. Hence, the phase
space for gauge and Higgs fields when interpreted as components of the
quantum noncommutative gauge field might not agree with their phase
space in conventional quantum theory.

These and other queries should be elucidated before we can claim
understanding of the meaning of quantization in NCG. Yet, by being
less demanding, we may perhaps realize whether the discussion of the
previous subsection is on the right track. If the quantum fluctuations
did not badly violate the constraints suggested by NCG, we might argue
that the latter would yield results that are not wide of the mark.
Indeed, it is common wisdom that the Standard Model will unavoidably
require modification at a certain energy scale. It may well happen
that a change in the quantization rules will also be needed around
that scale, yet it is probably sensible to assume that below such a
scale the standard quantization method makes a good approximation. We
shall then quantize our classical theory using the conventional method
and we shall regard it as an approximation to the ``true''
noncommutative quantum theory. Now that a quantization procedure has
been chosen, we are ready to analyze the quantum status of \eq{5.1}
and \eq{5.5}, henceforth called {\it NCG constraints}. We shall
discuss first whether the renormalization process that the formal
Green functions must undergo preserves the NCG constraints. The lack
of ultraviolet finiteness of our quantum field theory makes it
necessary to specify a renormalization scheme, so that $\mtop$, $m_H$
and $m_W$ can be given a unique meaning. The two most popular
renormalization schemes are the on-shell renormalization scheme
\cite{\Onshell} and the $\MSbar$ renormalization
scheme~\cite{\MSbarra}.

In the on-shell renormalization scheme, it would be quite natural to
decide that the NCG constraints in \eq{5.3} and \eq{5.5} should be
verified by the poles of the propagators. Hence, the previous
computations of~$x$ in terms of the top quark and $W$ boson masses
would effectively translate into predictions for the Higgs mass.

In the $\MSbar$ scheme, we shall denote $\mtop$, $m_H$ and $m_W$ by
$\mbar_{\rm top}(\mu)$, $\mbar_H(\mu)$ and $\mbar_W(\mu)$ to make
clear their dependence on the renormalization scale $\mu$. It is not
hard to see that, in the $\MSbar$ scheme, the NCG constraints are not
renormalization group invariant for any given value of the
parameter~$x$. Indeed, by taking into account the relations between
masses and coupling constants, which are given by
$$
\mbar_{\rm top}^2(\mu) = \thalf g_{\rm top}^2(\mu)\,\vbar^2(\mu),\quad
\mbar_H^2(\mu) = \thalf \la(\mu)\,\vbar^2(\mu),  \quad
\mbar_W^2(\mu) = \tquarter g_2^2(\mu)\,\vbar^2(\mu),
\eqno (5.6)
$$
one easily shows that the NCG constraints in the $\MSbar$ scheme are
respectively equivalent to the following equations:
$$
\eqalignno{
\a_{\rm top}(\mu) &= {2 - x \over 1 - x}\, \a_2(\mu),
& (5.7\,a) \cr
\a_H(\mu)
&= \biggl[3 - {3\over2}\Bigl({x^2 - 4x + 3\over x^2 - 7x + 10}\Bigr)
\biggr] \Bigl({2 - x \over 1 - x}\Bigr)\, \a_2(\mu).
& (5.7\,b) \cr}
$$
The symbol $\vbar(\mu)$ in \eq{5.6} denotes the VEV of the neutral
component of the Higgs field ($\approx 246\GeV$) in the $\MSbar$
scheme. The coefficients $g_{\rm top}(\mu)$, $g_2(\mu)$ and
$\lambda(\mu)$ are the top Yukawa, the $SU(2)$ coupling constant and
the Higgs quartic coupling constant, respectively. As usual,
$\a_{\rm top}(\mu) = g_{\rm top}^2(\mu)/4\pi$,
$\a_2(\mu) = g_2^2(\mu)/4\pi$ and $\a_H = \la/4\pi$. Now \eq{5.7} will
hold at any scale $\mu$ for a given value of $x$ only if the beta
functions $\b_{\rm top}$, $\b_H$ and $\b_2$, of $\a_{\rm top}$, $\a_H$
and $\a_2$, verify the following equations:
$$
\eqalign{
\b_{\rm top} &= {2 - x \over 1 - x}\, \b_2,
\cr
\b_H &= \biggl[ 3 - {3\over 2}
\Bigl({x^2 - 4x + 3 \over x^2 - 7x + 10}\Bigr) \biggr]
\Bigl({2 - x \over 1 - x}\Bigr)\, \b_2,
\cr}
$$
assuming that $x$ and $\mu$ are independent. We showed some time
ago~\cite{\Orpheus, \Persephone} that the one-loop beta functions in
the $\MSbar$ scheme do not verify \eq{5.6} for any value of~$x$;
therefore, at the one-loop level, the constraints are not preserved
under the renormalization flow, and in this sense, they do not
correspond to a hidden symmetry of the SM.

These one-loop beta functions are given~\cite{\Macha} by:
$$
\eqalignno{
4\pi \b_{\rm top}
&= \a_{\rm top}
(9\a_{\rm top} - 16\a_3 - {9\over 2}\a_2 - {17\over 10}\a_1),
\cr
4\pi \b_3 &= - 14 \a_3^2,
\cr
4\pi \b_2 &= - {19\over 3} \a_2^2,
& (5.8) \cr
4\pi \b_1 &= {41\over 5} \a_1^2,
\cr
4\pi \b_H
&= 6 \a_H^2 + 12 \a_H\a_{\rm top} - {9\over 5} \a_H\a_1 - 9 \a_H\a_2
+ {27\over 50} \a_1^2 + {9\over 5} \a_1\a_2 + {9 \over 2} \a_2^2
- 24 \a_{\rm top}^2.
\cr}
$$
where $\a_3(\mu) = g_3^2(\mu)/4\pi$, and
$\a_1(\mu) = {5\over 3} g_1^2(\mu)/4\pi$. (We are using an $SU(5)$
normalization, for which the Weinberg angle is defined through
$\a = \a_2 \sin^2\theta_W = {3\over 5} \a_1 \cos^2\theta_W$.)

The symbols $\b_1$, $\b_2$, $\b_3$ stand for the beta functions
corresponding to the gauge coupling constants of $U(1)$, $SU(2)$ and
$SU(3)$. We have neglected all fermionic Yukawas except the top. We
have also set to zero any Kobayashi--Maskawa phase, for their actual
values give rise to negligible corrections to our results. So far, in
our computations in the $\MSbar$ scheme, we have maintained the scalar
product coefficients $C_{\ell f}$, $C_{\ell q}$, $C_{qf}$ and $C_{qc}$
scale-independent, i.e., we have chosen the matrix $C$ in \eq{4.3}
independent of the renormalization scale~$\mu$. The question arises as
to whether this is too strong a requirement and, therefore, whether we
should allow the matrix $C$ to develop a dependence on the
renormalization scale. We should recall that it is the scalar product
in \eq{4.3} which yields the bosonic part of the renormalized
Lagrangian, the coefficients of the latter being renormalization scale
(even scheme) dependent in conventional quantum field theory. One
could then try to fix the $\mu$-dependence of the parameter~$x$ so
that \eq{5.7} became renormalization-group invariant. This idea ends
in failure since a clash occurs between \eq{5.7}, with a
$\mu$-dependent $x$, and \eq{5.8}. Solving for $x(\mu)$ the first
equation in \eq{5.7}, one obtains
$$
x(\mu)
= {\a_{\rm top}(\mu) - 2\a_2(\mu) \over \a_{\rm top}(\mu) - \a_2(\mu)}.
$$
Next, by plugging the previous equation into \eq{5.7$\,b$}, one fixes
the $\mu$-dependence of $\a_H(\mu)$. The beta function of $\a_H$ can
be the computed in terms of $\b_{\rm top}$, $\b_2$, $\a_{\rm top}$ and
$\a_2$. However, the beta function $\b_H$ thus obtained is not in
agreement with the renormalization group equations~\eq{5.8}.

In summary, we have shown that the bosonic part of the one-loop
renormalized SM Lagrangian cannot be recast at all scales into the
noncommutative form given by \eq{4.5}, even if we allow for scale
dependent~$C$'s.

\goodbreak

\subsection{5.3. Renormalization group results}

We have shown that the NCG constraints in \eq{5.2} and \eq{5.5} are
not preserved by quantum fluctuations in conventional quantum field
theory. What remains is to compute how badly these fluctuations
distort the NCG constraints. To give an accurate, and renormalization
group invariant, meaning to this statement, we shall proceed as
follows. First, we impose \eq{5.7} at a given scale, $\mu_0$.
Secondly, we compute $\a_{\rm top}$ and $\a_h$ at the $m_Z$ scale by
using the renormalization group equations \eq{5.8}. Finally, we work
out $\mbar_{\rm top}$ and $\mbar_H$ at $m_Z$ scale from \eq{5.6}. We
have repeated the process for several values of $x$ and $\mu_0$. The
results are presented in Tables 1--3. Before discussing them, let us
indicate that the $\a_i(\mu_0)$, $i = 1,2,3$ are obtained by running
the corresponding one-loop renormalization group equations in \eq{5.8}
from the $m_Z$ scale, with the initial conditions~\cite{\Langacker}:
$$
\eqalign{
\a_3(\mu = m_Z) &= 0.12,
\cr
\a_2(\mu = m_Z) &= 0.034,
\cr
\a_1(\mu = m_Z) &= 0.017,
\cr}
$$
up to the scale $\mu_0$. Notice that in our approximations these
coupling constants form a closed set under the renormalization group.

\begingroup
\parindent=0pt \leftskip=1cm \rightskip=1cm \baselineskip=11pt
\topinsert     %%_Table 1

\centerline{\vbox{\offinterlineskip
\def\tabrule{\noalign{\hrule}}
\def\hpt{height3pt&\omit&&\omit&&\omit&&\omit&&\omit&&\omit&&\omit&}
\halign{\vrule#&\strut\quad\hfil#\hfil\quad&\vrule width 1.2 pt#&
&\strut\quad\hfil#\hfil\quad&\vrule#\cr
\tabrule\hpt\cr
&$x$&& $\mu_0 = m_Z$&& $\mu_0 = 10^3$&& $\mu_0 = 10^4$&
&$\mu_0 = 10^5$&&$\mu_0 = 10^6$&& $\mu_0 = 10^7$&\cr
\hpt\cr
\noalign{\hrule height 1.2pt}
\hpt\cr
&$-1$&& 139&& 155&& 166&& 174&& 179&& 184&\cr
\hpt\cr \tabrule \hpt\cr
&$-0.75$&& 142&& 158&& 169&& 176&& 182&& 186&\cr
\hpt\cr \tabrule \hpt\cr
&$-0.50$&& 146&& 162&& 172&& 180&& 185&& 189&\cr
\hpt\cr \tabrule \hpt\cr
&$-0.25$&& 152&& 168&& 178&& 186&& 190&& 194&\cr
\hpt\cr \tabrule \hpt\cr
&0&& 161&& 176&& 186&& 193&& 197&& 199&\cr
\hpt\cr \tabrule \hpt\cr
&0.2&& 171&& 186&& 195&& 200&&\hfil&&\hfil&\cr
\hpt\cr \tabrule \hpt\cr
&0.4&& 185&& 200&& 207&& 211&&\hfil&&\hfill&\cr
\hpt\cr \tabrule \hpt\cr
&0.5&& 196&& 210&& 216&& 219&&\hfil&&\hfil&\cr
\hpt\cr \tabrule \hpt\cr
&0.7&& 236&& 244&&\hfil&&\hfil&&\hfil&&\hfil&\cr
\hpt\cr \tabrule \hpt\cr
&0.8&& 279&& 275&&\hfil&&\hfil&&\hfil&&\hfil&\cr
\hpt\cr \tabrule}}}
\vskip 12pt
{\bf Table 1:}
$\mbar_{\rm top}$ at the $m_Z$ scale as a function of $x$ and $\mu_0$.
Masses and energies are given in~GeV.

\endinsert     %%_Table 1
\endgroup

\begingroup
\parindent=0pt \leftskip=1cm \rightskip=1cm \baselineskip=11pt
\midinsert     %%_Table 2

\centerline{\vbox{\offinterlineskip
\def\tabrule{\noalign{\hrule}}
\def\hpt{height3pt&\omit&&\omit&&\omit&&\omit&&\omit&&\omit&&\omit&}
\halign{\vrule#&\strut\quad\hfil#\hfil\quad&\vrule width 1.2 pt#&
&\strut\quad\hfil#\hfil\quad&\vrule#\cr
\tabrule
\hpt\cr
&$x$&&$\mu_0 = m_{Z}$&&$\mu_0 = 10^3$&&$\mu_0 = 10^4$&
&$\mu_0 = 10^5$&&$\mu_0 = 10^6$&&$\mu_0 = 10^7$&\cr
\hpt\cr
\noalign{\hrule height 1.2pt}
\hpt\cr
&$-1$&& 212&& 201&& 195&& 192&& 191&& 191&\cr
\hpt\cr \tabrule \hpt\cr
&$-0.75$&& 220&& 207&& 200&& 197&& 196&& 195&\cr
\hpt\cr \tabrule \hpt\cr
&$-0.50$&& 228&& 214&& 206&& 202&& 200&& 200&\cr
\hpt\cr \tabrule \hpt\cr
&$-0.25$&& 240&& 223&& 214&& 209&& 207&& 206&\cr
\hpt\cr \tabrule \hpt\cr
&0&& 257&& 235&& 225&& 219&& 216&& 214&\cr
\hpt\cr \tabrule \hpt\cr
&0.2&& 276&& 250&& 236&& 229&&\hfil&&\hfil&\cr
\hpt\cr \tabrule \hpt\cr
&0.4&& 304&& 269&& 253&& 243&&\hfil&&\hfil&\cr
\hpt\cr \tabrule \hpt\cr
&0.5&& 325&& 284&& 264&& 253&&\hfil&&\hfil&\cr
\hpt\cr \tabrule \hpt\cr
&0.7&& 397&& 327&&\hfil&&\hfil&&\hfil&&\hfil&\cr
\hpt\cr \tabrule \hpt\cr
&0.8&& 472&& 366&&\hfil&&\hfil&&\hfil&&\hfil&\cr
\hpt\cr \tabrule}}}
\vskip 12pt
{\bf Table 2:}
$\mbar_H$ at the $m_Z$ scale as a function of $x$ and $\mu_0$. Masses
and energies are given in~GeV.

\endinsert     %%_Table 2
\endgroup

\begingroup
\parindent=0pt \leftskip=1cm \rightskip=1cm \baselineskip=11pt
\midinsert     %%_Table 3
\centerline{\vbox{\offinterlineskip
\def\tabrule{\noalign{\hrule}}
\def\hpt{height3pt&\omit&&\omit&&\omit&&\omit&&\omit&&\omit&&\omit&}
\halign{\vrule#&\strut\quad\hfil#\hfil\quad&\vrule width 1.2 pt#&
&\strut\quad\hfil#\hfil\quad&\vrule#\cr
\tabrule
\hpt\cr
&$x$&& $\mu_0 = m_Z$&& $\mu_0 = 10^3$&& $\mu_0 = 10^4$&
&$\mu_0 = 10^5$&& $\mu_0 = 10^6$&& $\mu_0 = 10^7$&\cr
\hpt\cr
\noalign{\hrule height 1.2pt}
\hpt\cr
&$-1$&& 1.53&& 1.30&& 1.17&& 1.10&& 1.07&& 1.04&\cr
\hpt\cr \tabrule \hpt\cr
&$-0.75$&& 1.55&& 1.31&& 1.18&& 1.12&& 1.08&& 1.04&\cr
\hpt\cr \tabrule \hpt\cr
&$-0.50$&& 1.56&& 1.32&& 1.120&& 1.12&& 1.08&& 1.06&\cr
\hpt\cr \tabrule \hpt\cr
&$-0.25$&& 1.58&& 1.33&& 1.20&& 1.12&& 1.09&& 1.06&\cr
\hpt\cr \tabrule \hpt\cr
&0&& 1.60&& 1.34&& 1.21&& 1.13&& 1.10&& 1.07&\cr
\hpt\cr \tabrule \hpt\cr
&0.2&& 1.61&& 1.34&& 1.21&& 1.15&&\hfil&&\hfil&\cr
\hpt\cr \tabrule \hpt\cr
&0.5&& 1.66&& 1.35&& 1.22&& 1.16&&\hfil&&\hfil&\cr
\hpt\cr \tabrule \hpt\cr
&0.7&& 1.68&& 1.34&&\hfil&&\hfil&&\hfil&&\hfil&\cr
\hpt\cr \tabrule \hpt\cr
&0.8&& 1.69&& 1.33&&\hfil&&\hfil&&\hfil&&\hfil&\cr
\hpt\cr \tabrule}}}
\vskip 12pt
{\bf Table 3:}
The ratio $\mbar_H/\mbar_{\rm top}$ at the $m_Z$ scale as a function
of $x$ and $\mu_0$. Masses and energies are given in~GeV.

\endinsert %%_Table 3
\endgroup

Our first comment applies to the blank boxes in the tables. Indeed,
for a given $x$, one cannot choose $\mu_0$ arbitrarily large, for
above a certain scale the solution to our renormalization group
equations diverge at $m_Z$. For instance, we cannot impose \eq{5.7}
much above $\mu_0 = 10^5\GeV$, if $x = 0.2$, since our solution to
\eq{5.8} will blow up at~$m_Z$. The highest energy one can reach
without meeting such problems is around $10^7\GeV$, but then $x$
cannot be much bigger than zero. A second comment has to do with the
stability of both $\mbar_{\rm top}(m_Z)$ and $\mbar_H(m_Z)$ as a
functions of $\mu_0$ and $x$. For any given $x$, both
$\mbar_{\rm top}(m_Z)$ and $\mbar_H(m_Z)$ vary between a minimum and a
maximum value as $\mu_0$ changes.

Let us define the error associated to a positive variable $w$ for
which $w_{\rm min} \leq w \leq w_{\rm max}$, as the ratio
$2 \x (w_{\rm max} - w_{\rm min})/(w_{\rm max} + w_{\rm min})$. Then,
for any given value of~$x$, the error associated to
$\mbar_{\rm top}(m_Z)$ is never much larger than $25$\%. This shows
that, for a given $x$, the ``predictions'' for $\mbar_{\rm top}(m_Z)$
and $\mbar_H(m_Z)$ depend weakly on the value of $\mu_0$, in
consonance with the logarithmic nature of the renormalization equation
corrections. In other words, the conventional quantization procedure
does not quite wipe out the NCG origin of our Lagrangian. So far, we
have kept $x$ fixed and allowed changes in $\mu_0$;%
\footnote*{keeping $\mu_0$ greater than $m_Z$; the alternative is not
being considered.}
if we also let $x$ vary between $-1$ and~1, a disaster occurs as
$x \to 1$, for the right side of \eq{5.7} becomes arbitrarily large.
We remind the reader that no predictions can be put forward in this
limit.

Let us now put a little physics, although of a somewhat speculative
nature, into our analysis. If we make the assumption that we cannot
trust the quantum field theory of the Standard Model above 1~TeV, as
currently made by some~\cite{\Veneziano}, then the physically
acceptable values for $x$ will be those that yield an
$\mbar_{\rm top}^{\rm exp}(m_Z)$ in agreement with its experimental
value, whatever the value of $\mu_0$ between $m_Z$ and $10^3\GeV$.
Taking $\mbar_{\rm top}^{\rm exp}(m_Z) \approx (175\pm 12)\GeV$ (this
is really the value of the pole mass of the top, but the error
involved in the approximation is probably less than $1$\%
\cite{\Kubo}), then from Table~1 we conclude that the acceptable $x$
values should verify $-0.25 < x < 0.4$. Armed with this constraint
on~$x$, and recalling the 1~TeV scale for $\mu_0$, we move down to
Table~2, and reach the conclusion that the values of $\mbar_H(m_Z)$
consistent with our assumptions lie in the interval $223{-}304\GeV$.
In summary, {\it classical} noncommutative geometry in combination
with the assumption that the conventional quantum Standard Model will
require some fixing at or below 1~TeV, leads to the following
``prediction'':
$$
223\GeV < \mbar_H(m_Z) < 304\GeV.
$$
This result is in good  agreement with the on-shell
interpretation of the NCG constraints \eq{5.2} and \eq{5.5};
inspection of Table~3 leads to the inequality quoted in the abstract.
The moral is that quantum fluctuations are likely to modify the
{\it prima facie\/} values suggested by the model for the Higgs mass
only in the direction of decreasing the lower bound. It seems unlikely
that this conclusion would be altered at the two-loop level.
Therefore, even if one were to allow $\mu_0$ as another free parameter
of the theory, the latter would still be falsifiable on experimental
grounds.

\medskip

\section{6. NCG and model building}

\subsection{6.1. Rigidity of NCG models}

Faced with the fresh outlook brought by noncommutative geometry to
particle physics, two philosophical attitudes were to be taken. One
was to regard the NCG notions as tools of a new freedom to theorize.
This was the path taken by the Z\"urich school in an impressive series
of papers~\cite{\CFFrohlich}.

Among other innovations, they chose to enlarge the space in which the
finite part Dirac operator lives, with extra degrees of freedom
associated to the vacuum expectation values of the several Higgs
fields they introduce. The interplay among the various aspects of
Connes' formulation is looser, and the consequent greater flexibility
in the choice of ingredients allowed them both to build several
unified gauge models and to deal with gravity and supersymmetry in
very interesting ways. The Z\"urich approach is perhaps more in line
with the original Kaluza--Klein motivation for NCG.

Other people welcomed instead the progressive tightening of Connes'
axiomatics as an opportunity to discriminate among the myriad of
Yang--Mills--Higgs models, with the more or less avowed aim of
demonstrating the world algebra to be minimal or preferred in some
sense. Whatever the final outcome of these efforts, it is indisputable
that familiarity with the Connes--Lott paradigm helps one to perceive
the seemingly arbitrary features of the Standard Model as engaging
subtleties of Nature.

In this paper, we decide to underline rather the rigidity of the NCG
requirements. As far as we know, the new scheme {\it en vogue\/} since
May 1995 \cite{\Tresttalk}, where charge conjugation symmetry plays
the prominent r\^ole, has not been treated in the Z\"urich manner.

Before proceeding, we wish to indicate that the very older scheme, in
which the tools of NCG are applied to just the sum of two copies of
spacetime, is alive and well \cite{\Wulkenhaar}. Also, among the
sources of the NCG approach to particle physics, we call attention to
the ideas of Dubois-Violette and coworkers~\cite{\DuboisVKM}.

Actually, in the orthodox NCG approach there are very strong
restrictions on gauge symmetries. Even leaving aside for the moment
Connes' geometrodynamical principle, gauge groups can only appear as
groups of unitary elements of algebras of matrices, and the left and
right actions of the group on fermions can only be a sum of copies of
the fundamental representation and its conjugate. Many models of the
Yang--Mills--Higgs type are ruled out by these principles, and/or
Poincar\'e duality (see below).

A very natural question, in that context, is to try to determine
whether by suitably choosing the $K$-cycle structure, the SM can be
obtained by the spontaneous symmetry breaking of semisimple algebras
containing the Eigenschaften algebra, in a similar way to the SSB of
$SU(3) \x SU(2)_L \x U(1)_Y$ to $SU(3) \x U(1)_{\rm em}$. (It is very
easy to see that simple algebras are ruled out, even if we allow
massive neutrinos to enter the picture. For instance, the main
candidate $SU(15)$, the unimodular unitary group of $M_{15}(\C)$,
breaks down to $SU(3) \x SU(12) \x U(1) \subset SU(15)$. The
fundamental representation would decompose as
${\bf 15} = ({\bf 3},{\bf 1},\la) \oplus ({\bf 1},{\bf 12},-\la/4)$,
implying that colour triplets are $SU(2)$ singlets.)

The question has been examined by Lizzi {\it et al\/}~\cite{\Neapolis}
under some additional natural assumptions. Their detailed analysis of
the possible semisimple algebras leads to the surprising conclusion
that no larger gauge groups are compatible with the noncommutative
structure of the SM.

\subsection{6.2. Poincar\'e duality for noncommutative manifolds}

Up to here, we have been somewhat cavalier with the concept of
noncommutative manifold, equating it more or less with a
noncommutative algebra. Actually, the concept of manifold is a subtle
one, both in commutative \cite{\Sullivan} and in noncommutative
geometry. Moreover, in the noncommutative case, the manifold depends
in an essential way on the particular representation of the algebra
one uses; and not every algebra can be regarded as a noncommutative
manifold. Since NCG models are expected to retain and extend the full
geometrical character of smooth manifolds beyond the commutative
setting, it is important to know precisely what a noncommutative
manifold is. This point is discussed at some length by Connes
\cite{\Book, \ConnesGrav, \RealNCG}, and his answer motivated the
construction of the colour sector in the original Connes--Lott models.

Algebraic tools that are not sensitive to continuous deformations tend
to lose some information about the point sets they describe; even so,
one tool that separates closed manifolds from manifolds with boundary
is Poincar\'e duality. For a compact Riemannian manifold of
dimension~$n$, this can be expressed as saying that the Hodge star
operator matches differential forms of degree~$k$ to those of degree
$(n - k)$ isomorphically. Alternatively, it says that the bilinear
pairing of forms
$$
(\a,\b) \longmapsto \int \a \w {\star \b}
\eqno (6.1)
$$
is {\it nondegenerate}. Since de~Rham cohomology is entirely given by
the Dirac $K$-cycle, we may expect this to have a nice formulation in
NCG, for more general $K$-cycles. What is perhaps not so obvious is
how to set up the duality so that it makes sense for zero-dimensional
(noncommutative) spaces, such as the finite matrix algebras we have
been dealing with here.

To proceed, we take our cue from the Chern character isomorphism that
links the de~Rham cohomology of a compact manifold to its $K$-theory
groups. (There are two of these, $K_0$ which is linked to even-degree
forms, and $K_1$ which is tied to odd-degree forms by the Chern
character; we shall consider only $K_0$.) For an involutive algebra
$\Al$, the abelian group $K_0(\Al)$ is generated by (classes of)
selfadjoint idempotent matrices ---projections--- with elements
in~$\Al$, modulo a suitable notion of equivalence; addition
corresponds to matrix direct sums. Thus for $\Al = \C$ or~$\HH$, we
get $K_0(\C) = \Z$ and $K_0(\HH) = \Z$, since the class of a
projection matrix is determined by its rank (or by its trace). For
$\Al = M_3(\C)$, say, we also get $K_0(\Al) = \Z$, where the generator
is the rank-one projection $e = \diag(1,0,0) \in M_3(\C)$. For the
algebra $\Al = \C \oplus \HH \oplus M_3(\C)$, the result is that
$K_0(\Al) = \Z \oplus \Z \oplus \Z$, with $1_\C$, $1_\HH$ and $e$
being independent generators.

The Chern character maps matrix projections over $\Al = C^\infty(M)$
to even-degree forms that are ``rational'' in the sense that the
pairing of forms \eq{6.1} yields rational values, when the integration
has been properly normalized. So Poincar\'e duality can be formulated
algebraically as a certain {\it nondegenerate\/} bilinear pairing on
$K_0(C^\infty(M))$ with values in~$\Q$. It turns out that the Chern
character makes perfect sense in noncommutative geometry (a long
story, told in Chapter~3 of~\cite{\Book}) and is given by a certain
supertrace of products of algebra elements $a$ and commutators
$[D,a]$. However, it is simpler and better to deal directly with the
$K$-theory picture. We are indebted to Daniel Testard~\cite{\Testard}
for explaining what follows to us.

The bilinear form on $K_0(\Al)$ depends on the chosen real $K$-cycle
$(\Al,\H,\D,\ga,J)$: to a pair of projections $(p,q)$ in $\Al$, or
more generally in $M_n(\Al)$ for some~$n$, it associates the index of
the Fredholm operator $e \D^+ e$ on the range space of the projector
$e = p JqJ^\7 = JqJ^\7 p$ on $\H^n$, if one writes
$\D = \pmatrix{0 & \D^- \cr \D^+ & 0 \cr}$ after splitting $\H$
according to chirality. For this index map to make sense in general,
the commutation conditions \eq{3.5$\,b,c$} are both required.

For our purposes, it is enough to know that for a noncommutative space
of dimension zero, the index formula given by the Chern character does
not involve the commutators $[\D,a]$ explicitly, but only the
chirality~$\ga$. In fact, the recipe for the pairing is
$$
(p,q) \longmapsto \Str(p J q J^\7) = \Tr(\ga p J q J^\7).
\eqno (6.2)
$$
This is symmetric since $J \ga = \ga J$ in dimension zero. To sum up,
Poincar\'e duality entails three conditions on a finite matrix algebra
$\Al$ with a real $K$-cycle:
$$
\eqalignno{
&[a, JbJ^\7] = 0, \sepword{for all} a,b \in \Al,
& (3.5\,b) \cr
&[[\D,a], JbJ^\7] = 0, \sepword{for all} a,b \in \Al,
& (3.5\,c) \cr
&(p,q) \longmapsto \Tr(\ga p J q J^\7) \qquad\hbox{is nondegenerate.}
& (6.3) \cr}
$$

We now compute this pairing for the finite-part $K$-cycle of the
Standard Model, with $\Al = \C \oplus \HH \oplus M_3(\C)$. The task is
to find the matrix with entries \eq{6.2}, where $p$ and $q$ are
replaced in turn by three generators of $K_0(\Al)$. Instead of the
obvious candidates $1_\C$, $1_\HH$ and $e$, let us take the following
generators: $p_1 = (-1_\C) \oplus e$, $p_2 = 1_\C \oplus 1_\HH$,
$p_3 = 1_\C$. In the case $p = q = p_1$, the representation \eq{4.1}
gives the following diagonal matrices:
$$
\eqalignno{
\ga &\mapsto (1,-1,-1)^N \oplus (1,1,-1,-1)^{3N} \oplus (1,-1,-1)^N
\oplus (1,1,-1,-1)^{3N},
\cr
p_1 &\mapsto (-1,0,0)^N \oplus (-1,-1,0,0)^{3N} \oplus (-1,-1,-1)^N
\oplus (e,e,e,e)^N,
\cr
J p_1 J^\7 &\mapsto (-1,-1,-1)^N \oplus (e,e,e,e)^N \oplus (-1,0,0)^N
\oplus (-1,-1,0,0)^{3N},
& (6.4) \cr}
$$
where, for instance, $(1,-1,-1)^N$ means $N$ diagonal blocks with
diagonal entries $1,-1,-1$. Thus $\Tr(\ga p_1 J p_1 J^\7) = -2N$.
Using the representations
$$
\eqalignno{
p_2 &\mapsto (1,1,1)^N \oplus (1,1,1,1)^{3N} \oplus (1,1,1)^N
\oplus (0,0,0,0)^{3N},
\cr
p_3 &\mapsto (1,0,0)^N \oplus (1,1,0,0)^{3N} \oplus (1,1,1)^N
\oplus (0,0,0,0)^{3N},
& (6.5) \cr}
$$
we quickly compute the matrix of the pairing ---with $(i,j)$-entry
$\Tr(\ga p_i J p_j J^\7)$--- to be
$$
2N \pmatrix{-1 & 0 & 0 \cr 0 & -1 & 0 \cr 0 & 0 & 1 \cr},
$$
so that \eq{6.2} is a {\it nonsingular\/} quadratic form.

All of that means that the finite part $K$-cycle of the SM satisfies
Poincar\'e duality, and so can legitimately be called a
``noncommutative 0-dimensional manifold''. The full SM is thus a
noncommutative 4-dimensional manifold.

This is no mere computational formality. Consider what
happens~\cite{\Testard} when one tries to modify the Standard Model to
allow for massive neutrinos. The diagonal blocks $(1,-1,-1)^N$,
$(-1,0,0)^N$, $(-1,-1,-1)^N$ in \eq{6.4} become $(1,1,-1,-1)^N$,
$(-1,-1,0,0)^N$ and $(-1,-1,-1,-1)^N$ respectively, to allow for the
presence of right neutrinos $\nu_R$. On recomputing the matrix of the
pairing, one gets
$$
2N \pmatrix{0 & -1 & -1 \cr -1 & 0 & 1 \cr -1 & 1 & 2 \cr},
$$
which is a {\it singular\/} matrix. Any possibility of matching the
algebra $\Al$ to a dual partner with this $K$-cycle has vanished. It
would seem, then, that the extension of the SM to include neutrinos
with Dirac masses is forbidden, if one were to accept that the SM is a
manifestation of an underlying noncommutative manifold. Nondegeneracy
of the Poincar\'e quadratic form seems to require some degree of
asymmetry between the lepton and quark sectors. If $u_R$ quarks were
absent, we would also get a singular intersection form.

\subsection{6.3. Poincar\'e duality and the colour sector}

The Poincar\'e duality condition may also be called upon to decide
whether the representation \eq{4.1} is, in a suitable sense, the only
possible one allowable in an NCG model if one starts from the
finite-part algebra $\Al = \C \oplus \HH \oplus M_3(\C)$.
Asquith~\cite{\Becca} has shown that the colour part of the algebra
must act vectorially, if one assumes the usual flavour action and
imposes Poincar\'e duality.

We specify that the flavour subrepresentation decomposes as before,
according to chirality:
$$
\pi^+(\la,q,m)
= \bordermatrix{&\scriptstyle{\ell R}&\scriptstyle{qR}
&\scriptstyle{\ell L}&\scriptstyle{qL}\cr
\scriptstyle{\ell R} & \la &&&\cr
\scriptstyle{qR}     && \La \ox 1_3 &&\cr
\scriptstyle{\ell L} &&& q &\cr
\scriptstyle{qL}     &&&& q \ox 1_3 \cr},
$$
where $\La = \pmatrix{\la & 0 \cr 0 & \bar\la \cr}$. The condition
$\D J = J\D$ splits $\D$ likewise, as in~\eq{3.6}. We may work only on
the particle space $\H^+$ if we define
$\bar\pi(b) := J \pi^-(b^*) J^\7$. The task is to determine the form
of~$\bar\pi$. We shall take $N = 1$ for the sake of the argument.

The duality conditions of \eq{3.5$\,b,c$} now say that each
$\bar\pi(b)$ commutes with every $\pi^+(a)$ and every $[D,\pi^+(a)]$,
for $a,b \in \Al$. The commutation relations
$[\pi^+(a),\bar\pi(b)] = 0$ imply that $\bar\pi(\la,q,m)$ depends only
on $(\la,m)$ and has the following block structure:
$$
\pmatrix{\bar\pi_R(\la,m) &\cr &\bar\pi_L(\la,m) \cr}
= \pmatrix{a_1(\la) &&&\cr	& 1_2 \ox B_1(\la,m) &&\cr
&& 1_2 \ox a_2(\la) &\cr &&& 1_2 \ox B_2(\la,m) \cr},
$$
where each $a_i(\la)$ is either $\la$ or $\bar\la$, and each
$B_i(\la,m)$ belongs to $M_3(\C)$. The quaternions drop out for the
following reason: since all $\bar\pi(0,q,0)$ and all $\pi^+(0,q',0)$
commute, we get $\bar\pi(0,q,0) = 0$ unless $q$ is real; and since
$\bar\pi(0,1,0) = \bar\pi(0,j,0)^4 = 0$ also, $\bar\pi(\la,q,m)$
cannot depend on~$q$. The terms $a_i(\la)$ cannot depend on~$m$ since
the simple algebra $M_3(\C)$ has only the trivial one-dimensional
representation.

Since $[D, \pi^+(0,1)]$ is of the form
$\pmatrix{0 & M^\7 \cr -M & 0 \cr}$ from~\eq{3.1}, the lower left
block of the matrix $[[D, \pi^+(0,1)], \bar\pi(\la,m)]$ is
$\bar\pi_L(\la,m) M - M \bar\pi_R(\la,m)$; the duality condition
\eq{3.5$\,c$} is satisfied if and only if this vanishes identically;
this yields two conditions
$$
m_\ell \bigl( a_1(\la) - a_2(\la) \bigr) = 0,  \qquad
M_q \ox \bigl( B_1(\la,m) - B_2(\la,m) \bigr) = 0,
$$
where $m_\ell$, $M_q$ are the lepton and quark blocks of the Yukawa
matrix $M$. Assuming that the Yukawa matrix is invertible, we conclude
that Poincar\'e duality forces $a_1 = a_2$ and $B_1 = B_2$, i.e., the
representation $\bar\pi$ commutes with chirality. One may therefore
conclude~\cite{\Becca} that the vectorial character of the strong
force is a consequence of Poincar\'e duality in NCG.

The option $B(\la,m) = \la 1_3$ is also ruled out, since it would
entail $\pi^+(0,0,e) = \bar\pi(0,0,e) = 0$, which makes the pairing
\eq{6.2} degenerate; thus $B(\la,m) = m$ (up to a complex
conjugation). Since $a(1) = 1$ and $B(1,e) = e$, the duality pairing
reduces to the previous calculation \eq{6.5}.

\subsection{6.4. Unimodularity and anomalies}

In subsection~4.5 we saw how the unimodularity condition reduces the
symmetry group from $U(1) \x SU(2)_L \x U(3)$ down to
$U(1) \x SU(2)_L \x SU(3)$, the latter in the SM representation. The
very same reduction is achieved by demanding that the fermions of our
model carry an anomaly-free representation embedded in the NCG
representation of $U(1) \x SU(2)_L \x U(3)$. Thus, anomaly
cancellation is equivalent to unimodularity as far as the NCG model at
hand is concerned.

Before imposing unimodularity, the symmetry group of the NCG model is
the unitary group of $\Al_F$, i.e., $U(1) \x SU(2)_L \x U(3)
\,\simeq\, U(1)_B \x U(1)_A \x SU(2)_L \x SU(3) =: G$. Of course, the
fermions carry not an arbitrary representation of this group but the
one coming from the action of~$\Al_F$ on~$\H_F$. {}From \eq{4.4} one
learns that our fermions $\bar e_L$, $\ell_L = (e_L,\nu_L)$,
$\bar d_L$, $\bar u_L$ and $q_L = (d_L,u_L)$, carry the following
representation of~$G$:
$$
(2 y,0,{\bf1},{\bf1}) \,\oplus\, (-y,0,{\bf2},{\bf1})
\,\oplus\, (y,k,{\bf1},\bar{\bf3}) \,\oplus\, (-y,k,{\bf1},\bar{\bf3})
\,\oplus\, (0,-k,{\bf2},{\bf3}).
\eqno (6.6)
$$
The symbols $y$ and $k$ denote the respective $U(1)_B$ and $U(1)_A$
charge scales, so they do not vanish. It is not difficult to show that
the representation \eq{6.6} is anomalous. Indeed, the
$U(1)_B [SU(2)]^2$ anomaly cancellation condition does not hold, since
$$
\tr Y_B \{T^a_{SU(2)}, T^b_{SU(2)}\} \propto -y
$$
never vanishes.

Since the group of unitaries $G$ is not a consistent symmetry group at
the quantum level, one is led to seek subgroups of $G$ whose
subrepresentations are anomaly free. In so doing, we take the $SU(2)$
doublet and $SU(3)$ triplet structures as given experimental facts and
do not tamper with them. We then consider only subgroups of the type
$U(1) \x SU(2)_L \x SU(3)$, where the representation of the $U(1)$
factor is obtained by imposing one of the following linear constraints
on the fields $B$ and $\widehat A := \tthird \tr A$ in~\eq{4.4}:
either $B = 0$, $\widehat A = M$, or
$$
\widehat A = \a B, \quad B = N.
$$
(However, the restriction of considering only these kinds of subgroups
\cite{\Chiron} can be removed, so the result that the SM corresponds
to the maximal anomaly-free subrepresentation of the unitary group of
the world algebra is unconditionally true.)

The first case gives rise to the following representation of
$U(1) \x SU(2)_L \x SU(3)$:
$$
(0,{\bf1},{\bf1}) \,\oplus\, (0,{\bf2},{\bf1})
\,\oplus\, (x,{\bf1},\bar{\bf3}) \,\oplus\, (x,{\bf1},\bar{\bf3})
\,\oplus\, (-x,{\bf2},{\bf3}),
$$
where $x$ stands for the $U(1)$ charge scale. Now, this representation
is anomalous since $\tr Y_M \{T^a_{SU(2)}, T^b_{SU(2)}\} \propto -x$
never vanishes. In the second case, we have a family of
representations indexed by the parameter~$\a$:
$$
(2x,{\bf1},{\bf1}) \,\oplus\, (-x,{\bf2},{\bf1})
\,\oplus\, ((1+\a)x,{\bf1},\bar{\bf3})
\,\oplus\, ((\a-1)x,{\bf1},\bar{\bf3})\,\oplus\, (-\a x,{\bf2},{\bf3}).
$$
Only one member of this family is anomaly free. Indeed, the anomaly
cancellation condition $\tr Y_N^3 = 6 + 18\a = 0$ uniquely fixes $\a$
to $-\tthird$ (the unimodularity constraint!), yielding the following
representation of $U(1) \x SU(2)_L \x SU(3)$:
$$
(2x,{\bf1},{\bf1}) \,\oplus\, (-x,{\bf2},{\bf1})
\,\oplus\, (\tfrac 23 x,{\bf1},\bar{\bf3})
\,\oplus\, (-\tfrac 43 x,{\bf1},\bar{\bf3})
\,\oplus\, (\tthird x,{\bf2},{\bf3}).
$$
This is precisely the Standard Model representation.

Equivalence between anomaly cancellation and unimodularity is thus
established with\-in the context of the NCG model under study. This
equivalence does not hold for all NCG field theory
models~\cite{\Pris}.

It has long been known \cite{\Geng, \Ramond} that the anomaly
cancellation conditions almost determine the fermionic hypercharges
within the framework of the conventional formulation of the SM. Yet if
the mixed anomaly \cite{\LuisAG} involving two gravitons is not
considered, one hypercharge remains completely arbitrary~\cite{\Geng}.
On the other hand, we have here that the representation \eq{6.6} of
$G$ satisfies $\tr Y_B = 0$ and $\tr Y_A = 0$; hence, there is no
anomaly involving either the $U(1)_B$ current and two gravitons or the
$U(1)_A$ current and two gravitons. (Also, the representation \eq{6.6}
of $G$ has neither a $U(1)_B [SU(3)]^2$ nor a $U(1)_A [SU(3)]^2$
anomaly.) The fact that in the NCG reconstruction of the SM the mixed
gravitational anomaly is absent {\it ab initio\/} may be hinting at a
deep relationship between quantum physics and gravitation in the
noncommutative framework. It is tempting as well, in view of the
preceding, to interpret anomalies as obstructions to the existence of
noncommutative manifolds.

\section{7. Conclusions, or what NCG does (and does not do) for us}

We summarize the several quirky features of the SM, as a series of
questions and answers.

\smallskip
\item{Q1:}
The Higgs sector is introduced by hand.

\smallskip
\item{A:}
The Higgs is interpreted geometrically as a gauge boson in
Connes--Lott theory, and is unified with the other gauge bosons in a
unique noncommutative gauge field.

\smallskip
\item{Q2:}
The link between parity violation and the symmetry breaking sector
remains mysterious in the Standard Model.

\smallskip
\item{A:}
In Connes--Lott theory, chirality of the electroweak sector is
necessary for the Higgs sector to occur; moreover, Poincar\'e duality,
given the electroweak representation, forces the colour representation
to be non-chiral.

\smallskip
\item{Q3:}
There is no explanation for the observed number of fermionic
generations.

\smallskip
\item{A:}
For the Higgs mechanism to work, NCG requires the number of fermion
generations to be at least two; and with the present value of the mass
of the top, the model considered in this paper gives an upper bound of
five generations.

\smallskip
\item{Q4:}
The choice of gauge groups and hypercharge assignments seems
arbitrary.

\smallskip
\item{A:}
The choice of possible gauge groups is very much restricted in NCG,
and the hypercharge assignments are explained either by anomaly
cancellation or by Connes' geometrodynamical principle.

\smallskip
\item{Q5:}
There is an apparent juxtaposition of gauged and non-gauged
interaction sectors.

\smallskip
\item{A:}
The Yukawa interaction is gauged.

\smallskip
\item{Q6:}
There is no explanation for the huge span of fermionic masses.

\smallskip
\item{A:}
Noncommutative geometry cannot explain that, because it starts from
there as its basic datum. Nevertheless, it gives a rationale for the
absence of right-handed neutrinos.
\item{}
The model considered in this paper does have something to say about
the mass of the $W$~boson ---thus becoming a retrodiction on the mass
of the top--- and about the mass of the Higgs particle: it suggests
that the top mass should be {\it larger\/} than $\sqrt 3\,m_W$, and
that the Higgs mass should be {\it smaller\/} than $\sqrt 3\,\mtop$.
The first~3 corresponds to the number of particle species, and the
second~3 comes from colour~\cite{\Ramans}.

\section{Acknowledgments}

We are very grateful to E.~Alvarez for decisive help with Section~5,
and for discussions. We thank R.~Coquereaux for a careful reading of a
draft of the manuscript. We are also grateful for enlightening
discussions with A.~Connes, M.~J. Herrero, B.~Iochum, W.~Kalau,
D.~Kastler, G.~Landi, F.~Lizzi, G.~Mangano, G.~Miele, J.~Plass,
I.~Pris, M.~Quir\'os, A.~Rivero, F.~Scheck, T.~Sch\"ucker, G.~Sparano,
D.~Testard, M.~Walze and J.-M.~Warzecha. We also thank the referee,
who pointed out a mistake in the first version of the manuscript.

This paper grew out of an effort to understand the pioneering work of
D.~Kastler and T.~Sch\"ucker; if it shows, we shall be glad.

The support of CICYT of Spain and the Vicerrector{\'\i}a de
Investigaci\'on of UCR are acknowledged.

\section{References}

\frenchspacing

\refno\ConnesEssay.
A. Connes, in {\it The Interface of Mathematics and Particle Physics},
D. G. Quillen, G. B. Segal and S. T. Tsou, eds. (Clarendon Press,
Oxford, 1990)

\refno\ConnesLott.
A. Connes and J. Lott, Nucl. Phys. B (Proc. Suppl.) 18 (1990) 29

\refno\Book.
A. Connes, Noncommutative Geometry (Academic Press, London, 1994)

\refno\MariaJose.
M. J. Herrero, preprint hep-ph/9601286, Madrid, 1996

\refno\ConnesGrav.
A. Connes, preprint hep-th/9603053, IHES, 1996

\refno\Sirius.
J. C. V\'arilly and J. M. Gracia-Bond{\'\i}a, J. Geom. Phys. 12 (1993)
223

\refno\Blackone.
F. Abe {\it et al} (CDF collaboration), Phys. Rev. Lett. 74 (1995)
2626

\refno\Blacktwo.
S. Abachi {\it et al\/} (D0 collaboration), Phys. Rev. Lett. 74 (1995)
2632

\refno\Orpheus.
E. Alvarez, J.~M. Gracia-Bond{\'\i}a and C.~P. Mart{\'\i}n, Phys.
Lett. B306 (1993) 55

\refno\Persephone.
E. Alvarez, J.~M. Gracia-Bond{\'\i}a and C.~P. Mart{\'\i}n, Phys.
Lett. B329 (1994) 259

\refno\Tresttalk.
A. Connes, talks given at the Conference on Noncommutative Geometry
and its Applications, \Trest, Czech Republic, May 1995

\refno\RealNCG.
A. Connes, J. Math. Phys. 36 (1995) 6194

\refno\Proteus.
J.~M. Gracia-Bond{\'\i}a, Phys. Lett. B351 (1995) 510

\refno\ArtstateIV.
D. Kastler and T. Sch\"ucker, Rev. Math. Phys. 8 (1996) 205

\refno\SchZyl.
T. Sch\"ucker and J.-M. Zylinski, J. Geom. Phys. 16 (1995) 207

\refno\IocSch.
B. Iochum and T. Sch\"ucker, Commun. Math. Phys. 178 (1996) 1

\refno\Neapolis.
F. Lizzi, G. Mangano, G. Miele and G. Sparano, Mod. Phys. Lett. A11
(1996) 2561

\refno\Testard.
D. Testard, unpublished manuscript, 1995

\refno\Chiron.
E. Alvarez, J.~M. Gracia-Bond{\'\i}a and C.~P. Mart{\'\i}n, Phys.
Lett. B364 (1995) 33

\refno\Diracm.
P. A. M. Dirac, The Principles of Quantum Mechanics, 4th edition
(Clarendon Press, Oxford, 1958)

\refno\ChengLi.
T.-P. Cheng and L.-F. Li, Gauge theory of elementary particle physics
(Clarendon Press, Oxford, 1988)

\refno\Donoghue.
J. F. Donoghue, E. Golowich and B. R. Holstein, Dynamics of the
Standard Model (Cambridge University Press, Cambridge, 1992)

\refno\Nakahara.
M. Nakahara, Geometry, Topology and Physics (Adam Hilger, Bristol,
1990)

\refno\Witten.
E. Witten, Nucl. Phys. B186 (1981) 412

\refno\KaluzaK.
T. Appelquist, A. Chodos and P. G. O. Freund, Modern Kaluza--Klein
theories, (Addison-Wesley, Reading, 1987)

\refno\CammaCoq.
G. Cammarata and R. Coquereaux, preprint hep-th/9505192, Luminy, 1995

\refno\Peskin.
M. E. Peskin and D. V. Schroeder, An Introduction to Quantum Field
Theory (Addi\-son--Wesley, Reading, 1995)

\refno\Marshak.
R.~E. Marshak, Conceptual Foundations of Modern Particle Physics
(World Scientific, Singapore, 1993)

\refno\Belinda.
J.~M. Gracia-Bond{\'\i}a, in {\it A Gift of Prophecy, Essays in
Celebration of the life of Robert Eugene Marshak}, E.~C.~G. Sudarshan,
ed. (World Scientific, Singapore, 1995)

\refno\ConnAction.
A. Connes, Commun. Math. Phys. 117 (1988) 673

\refno\Plassthesis.
J. Plass,
%%_Verallgemeinerte_Differentialformen_als_Zugang_zur
%%_nichtkommutativen_Differentialgeometrie,
Diplomarbeit, Universit\"at Mainz, 1993

\refno\KPPW.
W. Kalau, N. A. Papadopoulos, J. Plass and J.-M. Warzecha, J. Geom.
Phys. 16 (1995) 149

\refno\CiprianiGS.
F. Cipriani, D. Guido and S. Scarlatti, J. Operator Theory 35 (1996)
179

\refno\KastlerEH.
D. Kastler, Commun. Math. Phys. 166 (1995) 633

\refno\WolfMarkus.
W. Kalau and M. Walze, J. Geom. Phys. 16 (1995) 327

\refno\Connesphone.
A. Connes, private communication

\refno\ChamCon.
A. Chamseddine and A. Connes, preprint hep-th/9606001, IHES, 1996

\refno\Kalau.
W. Kalau, J. Geom. Phys. 18 (1996) 349

\refno\CIKS.
L. Carminati, B. Iochum, D. Kastler and T. Sch\"ucker,
preprint hep-th/9612228, Luminy, 1996

\refno\DoubleTrouble.
F. Lizzi, G. Mangano, G. Miele and G. Sparano,
preprint hep-th/9610035, Napoli and Oxford, 1996

\refno\Coque.
R. Coquereaux, G. Esposito-Far\`ese and F. Scheck, Int. J. Mod. Phys.
A7 (1992) 6555

\refno\Mainz.
R. H\"au{\ss}ling and F. Scheck, Phys. Lett. B336 (1994) 477

\refno\Fuzzy.
L. Carminati, B. Iochum and T. Sch\"ucker, preprint hep-th/9604169,
Luminy, 1996

\refno\Particledata.
R. M. Barnett et al., Phys. Rev. D54 (1996) 1

\refno\Onshell.
W. F. L. Hollick, Fortschr. Phys. 78 (1990) 165

\refno\MSbarra.
S. Franchiotti and A. Sirlin, Phys. Rev. D41 (1990) 319 and references
therein

\refno\Macha.
M. E. Machacek and M. T. Vaughn, Nucl. Phys. B222 (1983) 83;
Nucl. Phys. B236 (1984) 221; Nucl. Phys. B249 (1985) 70

\refno\Langacker.
P. Langacker, ed., Precision Tests of the Standard Electroweak Model
(World Scientific, Singapore, 1995)

\refno\Veneziano.
G. Veneziano, summary talk at the 9th International Workshop on
Supersymmetry and Unification of Fundamental Forces, Palaiseau, 13--19
May 1995, CERN--TH--95--332

\refno\Kubo.
J. Kubo, Phys. Lett. B 472 (1991) 262

\refno\CFFrohlich.
A. H. Chamseddine, G. Felder and J. Fr\"ohlich, Nucl. Phys. B395
(1993) 672;
A. H. Chamseddine and J. Fr\"ohlich, Phys. Lett. B314 (1993) 308;
A. H. Chamseddine, J. Fr\"ohlich and O. Grandjean, J. Math. Phys. 36
(1995) 6255, inter alia

\refno\Wulkenhaar.
R. Wulkenhaar, J. Math. Phys. 37 (1996) 3797

\refno\DuboisVKM.
M. Dubois-Violette, R. Kerner and J. Madore,
Phys. Lett. B217 (1989) 485; J. Math. Phys. 31 (1990) 316, 323,
inter alia

\refno\Sullivan.
D. Sullivan, Lecture Notes in Maths. 197 (Springer, Berlin, 1971)

\refno\Becca.
R. E. Asquith, Phys. Lett. B366 (1996) 220

\refno\Pris.
I. Pris and T. Sch\"ucker, preprint hep-th/9604115, Luminy, 1996

\refno\Geng.
C. Geng and R. Marshak, Phys. Rev. D39 (1989) 693

\refno\Ramond.
J. Minahan, P. Ramond and R. C. Warner, Phys. Rev. D41 (1990) 715

\refno\LuisAG.
L. Alvarez-Gaum\'e and E. Witten, Nucl. Phys. B234 (1983) 269

\refno\Ramans.
A. C. Clarke, Rendezvous with Rama (Bantam, New York, 1973):
``the Ramans do everything in threes''

\bye